\documentclass[pra, preprint, superscriptaddress]{revtex4}
\usepackage{graphicx}
\usepackage{dcolumn}
\usepackage{amsmath}
\usepackage{amssymb}
\usepackage{color}
\usepackage{float}
\usepackage{epstopdf}

\def\beq{\begin{equation}}
\def\eeq{\end{equation}}
\def\be{\begin{equation}}
\def\ee{\end{equation}}
\def\bea{\begin{eqnarray}}
\def\eea{\end{eqnarray}}
\def\eq#1{(\ref{#1})}
\def\simg{\,\hbox{\kern.1em \lower.6ex \hbox{$\sim$} \kern-1.12em
          \raise.6ex \hbox{$>$} }}
\def\siml{\,\hbox{\kern.1em \lower.6ex \hbox{$\sim$} \kern-1.12em
          \raise.6ex \hbox{$<$} }}
\newcommand{\Table}[4]{
\begin{table}[H]\begin{center}{#3}
\parbox{#2cm}{
\caption[table]{\renewcommand{\baselinestretch}{0.8} \small
                                           \hspace{-0.3truecm}#4}
\label{#1}}
\end{center}
\end{table}
}

\begin{document}

\title{Semiclassical analysis of distinct square partitions}

\author{M. V. N. Murthy}
\affiliation{Institute of Mathematical Sciences, Chennai, 600 113 India}
\author{Matthias Brack\footnote{e-mail address: matthias.brack@ur.de}}
\affiliation{Institute for Theoretical Physics, University of Regensburg,
D-93040 Regensburg, Germany}
\author{Rajat K. Bhaduri}
\affiliation{Department of Physics and Astronomy,  
McMaster University, Hamilton, Canada L9H 6T6}
\author{Johann Bartel}
\affiliation{IPHC, Physique Th\'eorique,
Universit\'e de Strasbourg, F-67037 Strasbourg, France}

\date{\today}

\begin{abstract}

We study the number $P(n)$ of partitions of an integer $n$ into sums of distinct 
squares and derive an integral representation of the function $P(n)$. Using 
semiclassical and quantum statistical methods, we determine its asymptotic 
average part $P_{as}(n)$, deriving higher-order contributions to the known 
leading-order expression [M. Tran {\it et al.}, Ann.\ Phys.\ (N.Y.) {\bf 311}, 
204 (2004)], which yield a faster convergence to the average values of the exact 
$P(n)$. From the Fourier spectrum of $P(n)$ we obtain hints that integer-valued
frequencies belonging to the smallest Pythagorean triples $(m,p,q)$ of integers 
with $m^2+p^2=q^2$ play an important role in the oscillations of $P(n)$. Finally 
we analyze the oscillating part $\delta P(n)=P(n)-P_{as}(n)$ in the spirit of 
semiclassical periodic orbit theory [M. Brack and R. K. Bhaduri: {\it 
Semiclassical Physics} (Bolder, Westview Press, 2003)]. A semiclassical trace 
formula is derived which accurately reproduces the exact $\delta P(n)$ for $n\simg 
500$ using 10 pairs of `orbits'. For $n\simg 4000$ only two pairs of orbits with the 
frequencies 4 and 5 -- belonging to the lowest Pythagorean triple (3,4,5) -- are
relevant and create the prominent beating pattern in the oscillations. For $n\simg 
100,000$ the beat fades away and the oscillations are given by just one pair of 
orbits with frequency 4.

\end{abstract}

\maketitle
     
\section{Introduction}
\label{secintro}

Consider a one-dimensional trap with an integer-valued quantum spectrum.
The problem of counting the number of excited states at a given energy $E$ 
is the same as writing an integer as a sum of its parts. For example, 
partitioning the excitation energy in a harmonic spectrum is the same as 
partitioning an integer into a sum of other integers. Similarly, 
partitioning the excitation energy in a one-dimensional box with infinitely 
steep reflecting walls corresponds to partitioning an integer into sums of 
squares of integers \cite{tran}. This connection leads to a link between number 
theory and statistical mechanics. A partition may be fully unrestricted, allowing 
all possible sums with repetitions (bosonic), or allowing only distinct entries 
in the summands (fermionic). It may also be restricted by allowing only a fixed 
number $N$ of summands in each partition.

Asymptotic formulae for large $n$ are known following the work of Hardy and
Ramanujan \cite{hr} and are found in many text books (see, e.g., \cite{apostol}). 
This year marks the centenary of the publication of the famous paper by Hardy and 
Ramanujan \cite{hr}, and it is appropriate to study the problem further. In 
particular, it is worthwhile to examine the number $P(n)$ of partitions of an 
integer $n$ into sums of distinct squares (hereafter called F2 partitions), which 
contain some unique features. 
As was pointed out in Ref.\ \cite{tran}, the exact function $P(n)$ for distinct 
square partitions exhibits pronounced oscillations with a beat-like structure 
when the points are joined by a continuous curve, as shown in Fig.\ \ref{figP500}
(cf. also Figs.\ \ref{figbes1} and \ref{figbes2} below).
\begin{figure}[h]
\centering
\vspace*{-0.8cm}
\hspace*{-1.5cm}\includegraphics[width=12cm,angle=0]{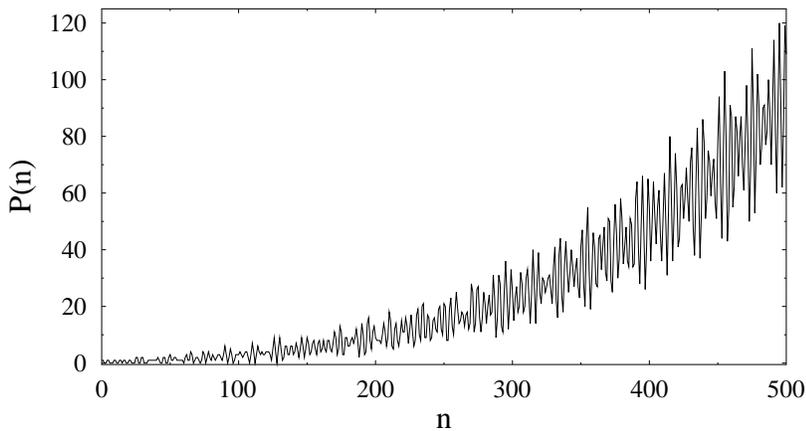}
\vspace*{-0.6cm}
\caption{$P(n)$ of the F2 partitions, shown in the low-$n$ region. Note that
$P(n)$ is given for integer values of $n$. Here we have joined the points
by a continuous curve to emphasize the beat structure.}
\label{figP500}
\end{figure} 

It is interesting to note that the beat structure eventually fades out for 
$n\simg 100,000$ (as shown in Sec.\ \ref{sectrf}), while the oscillations 
persist even as $n\rightarrow \infty$. Where are these regular oscillations 
coming from? Consider an integer $n$ that is a sum of squares: $n=m^2+p^2$. 
If $n$ itself is a square: $n=q^2$, then the three numbers $m,p,q$ form what 
is commonly called a Pythagorean triple (PT) of integers $(m,p,q)$ with 
$m^2+p^2=q^2$. Such triples can only occur in square partitions, since 
Fermat's last theorem \cite{wiles} asserts that only squares of integers 
may be written as sums of two (or more) other squares. Since an increasing
number of such triples will occur in the F2 partitions with increasing $n$, 
it is quite plausible that they reflect themselves in the oscillatory 
behaviour of the function $P(n)$.

In semiclassical physics \cite{gutz,book}, oscillatory behavior in the 
quantum-mechanical density of state of a dynamical system are described by a 
superposition of the periodic orbits of the corresponding classical system. 
The quantitative link between the quantum and classical description is called 
a `trace formula' \cite{gutz}. In many examples it has been shown \cite{book} 
that the gross features of quantum oscillations may be interpreted in 
terms of the shortest periodic orbits of the system, whose interference
often leads to beating patterns. One of the goals of the present paper is 
to find a semiclassical trace formula for the oscillations in $P(n)$. 

As mentioned above, F2 partitions express an integer $n$ as a sum of distinct 
squares. This is analogous to the distribution of the total energy $E= n = 
m_1^2+m_2^2+\dots$ amongst unrestricted numbers of Fermions in a one-dimensional 
box whose spectrum is given by the squares of integer quantum numbers $m_i$ 
(when all scales are set to unity). Thus we can map the F2 partitions onto a 
dynamical system in which fermions move in a one-dimensional box. This 
allows us to use the language of dynamical systems, with the notion of `energy' 
and its conjugate variable defined in Fourier space, the (time) `period'. The 
inverse of the period is then the `frequency' of the classical `orbit' in the 
dynamical system (though all these quantities are dimensionless in the present 
case). We will demonstrate that the orbits with the frequencies 4 and 5 that 
occur in the smallest PT (3,4,5) are sufficient to reproduce the beating 
oscillations in $\delta P(n)= P(n)-P_{as}(n)$ of the F2 partitions for $n\simg 
4000$. For smaller $n$, more orbits -- amongst them with frequencies 12 and 13 
contained in the PT (5,12,13) -- are needed to reproduce $\delta P(n)$, while 
in the asymptotic domain $n\simg 100,000$, where the beat fades away, the orbits 
with frequency 4 alone reproduce the correct oscillations.

Our paper is structured as follows. Section \ref{secbasic} contains our basic 
definitions and some formal results. Amongst them is an integral representation 
of $P(n)$ that later serves as a starting point for our semiclassical studies. 
In Sec.\ \ref{secasy} we focus on the asymptotic smooth part $P_{as}(n)$. We use 
the stationary-phase method for not only re-deriving the leading-order asymptotic 
expression $P^{(0)}_{as}(n)$ given in \cite{tran}, but also to find saddle-point 
corrections which lead to a faster convergence of $P_{as}(n)$ to the average of 
the exact $P(n)$. We show that the limit $P(n)/P_{as}(n)\rightarrow 1$ is 
practically reached for $n \sim 10,000$.

In Sec.\ \ref{secft} we present the Fourier spectrum of $P(n)$. We classify the 
leading peaks of the F2 spectrum into successive generations with decreasing 
intensities and observe a dominant presence of pairs of Fourier peaks whose 
frequencies are related to PTs. This suggests that the main oscillations in 
$P(n)$ may, indeed, be governed by the smallest PTs.

Finally, in Sec.\ \ref{sectrf}, we focus on the oscillating part $\delta P(n)$. 
Using stationary-phase integration over a few leading saddles in the complex 
$\beta$ plane that correspond to the dominating Fourier peaks, we derive a 
semiclassical trace formula for $\delta P(n)$. The results converge very 
fast upon including successive generations of orbits. In fact, the exact 
$\delta P(n)$ is reproduced by the trace formula already for $n \simg 4000$ 
using only the orbits with the frequencies 4 and 5 appearing in the PT (3,4,5). 
The rapid oscillations in $\delta P(n)$ have roughly the period of the orbit 
with the largest amplitude (frequency 4), while the period of the beating 
amplitude is given by the inverse difference 20 of their frequencies.

In Appendix \ref{secspcor} we explain details of the saddle-point corrections 
for the smooth part of the partition density, and in Appendix \ref{secairy} 
we illustrate the stationary-phase integration method in the complex plane
as a tool for evaluating asymptotic oscillations for the model case of 
the Airy function. 


\section{Basic definitions and formal results}
\label{secbasic}

\subsection{Unrestricted square partitions}
\label{ssecusp}

The function $P(n)$ counts the number of ways in which a given integer $n$
can be written as a sum of distinct squares of positive integers $m_i$:
\beq
n = \sum_{i=1}^{I_n} m_i^2\,, \qquad m_i\neq m_j\; \hbox{ for }\; i\neq j\,.
\label{partsum}
\eeq

Hereby the number $I_n$ of summands is not specified. It may start from 
$I_n=1$, in which case $m_1$ is the largest integer $\leq\!\sqrt{n}$. 
The highest $I_n$ is limited by the value of $n$ itself and may be found
from summing the lowest $I_n$ distinct squares, so that
\beq
\sum_{i=1}^{I_n} i^2=I_n(I_n+1)(2I_n+1)/6\leq n\,.
\eeq
This leads for large $n$ to the upper limit
\beq
I_n \leq (3n)^{1/3}-1/2 +{\cal O}(n^{-1/3})\,.
\label{imax}
\eeq
Each particular sum \eq{partsum} is called a `partition' of $n$ into squares. 
The word `distinct' implies that all $m_i$ within each partition must be 
different. This is analogous to the distribution of single-particle energies 
among fermions in statistical mechanics at a given total energy. We therefore
use the acronym `F2' for these partitions, where `F' stands for fermionic and 
`2' for squares. The word `unrestricted' specifies the fact that the number 
$I_n$ in \eq{partsum} is not fixed. We define $P(0)=1$ and, trivially, one 
sees that $P(1)=1$. The infinite series of numbers given by $P(n)$ for 
$n=0,1,2,\dots$ is called the series A033461 in the on-line encyclopedia of 
integer sequences (OEIS) \cite{oeis}. Its first ten members are 1, 1, 0, 0, 
1, 1, 0, 0, 0, 1.
  
The partition function $Z(\beta)$ for the unrestricted F2 partitions was given 
in \cite{tran} in several forms. We may obtain it as a generating function, 
which for any given partition $P(n)$ is defined as
\beq
Z(\beta)=\sum_{n=0}^\infty P(n)\,e^{-n\beta}.
\label{Zgen}
\eeq
For the F2 partitions it becomes (cf. Table 14.1 of \cite{apostol})
\beq
Z(\beta)=\prod_{m=1}^\infty [1+e^{-m^2\beta}],
\label{Zprod}
\eeq
which was also used to generate our data base for the $P(n)$ up to $n=160,000$.

In the following we write the complex variable $\beta$ as
\beq
\beta=x+i\tau\,,\qquad (x,\tau\in{\rm R})\,,
\eeq
where $x$ and $\tau$ are dimensionless real variables. Note that \eq{Zprod} can 
be viewed as a fermionic canonical grand partition function with chemical 
potential $\mu=0$. Therefore there is no constraint on the average particle number 
$N$ which may go up to infinity. Its single-particle spectrum is given by integer 
squares, as for a particle in an infinite square box (with dimensionless energy 
and spatial units).

The inverse Laplace transform of $Z(\beta)$ yields the partition density $g(E)$:
\beq
g(E) = {\cal L}^{-1}_E [Z(\beta)] 
     = \frac{1}{2\pi i}\int_C Z(\beta)\,e^{E\beta}\,{\rm d}\beta\,.
\label{LapiP}
\eeq
Here $E$ is a dimensionless real variable. We choose here the symbol $E$ because 
of its relation to the energy in the context of statistical physics, where $g(E)$ 
is the level density (or density of states) of a system of independent particles. 

The contour $C$ in \eq{LapiP} runs parallel to the imaginary axis $\tau$ with a 
real part $x=\epsilon>0$, so that we may write
\beq
g(E) = \frac{1}{2\pi} \int_{-\infty}^{+\infty} 
       Z(\epsilon+i\tau)\,e^{E(\epsilon+i\tau)}\,{\rm d}\tau \qquad (\epsilon>0)\,.
\eeq
The (two-sided) Laplace transform of $g(E)$ gives back the partition 
function $Z(\beta)$:
\beq
{\cal L}_\beta [g(E)] = \int_{-\infty}^{+\infty}g(E)\,e^{-E\beta}\,{\rm d}E
                      = Z(\beta)\,.
\label{LapP}
\eeq
Using the form \eq{Zgen}, the density of F2 partitions is immediately found to be
\beq
g(E) = \sum_{n=0}^\infty P(n)\, \delta(E-n)\,,
\label{gofedel}
\eeq
where $\delta(E-n)$ is the Dirac delta function peaked at $E=n$.

In order to recover the $P(n)$ from the partition density \eq{gofedel}, we just 
have to integrate it over a small interval around $E=n$:
\beq
P(n) = \int_{n-a}^{n+a} g(E)\,{\rm d}E\,, \qquad  (0 < a < 1)\,.
\label{Pn}
\eeq
If we choose $a=1/2$, Eq.\ \eq{Pn} corresponds to an averaging of $g(E)$ over a unit 
interval $\Delta n = 1$ around $n$. Averaging $g(E)$ over larger intervals $\Delta E$ 
therefore corresponds to averaging the $P(n)$ over some larger interval $\Delta n$:
\beq
\langle\, g(E)\, \rangle_{\Delta E} \; \sim \; \langle\, P(n)\, \rangle_{\Delta n}\,.
\label{averages}
\eeq
This property will be used in Sec.\ \ref{secasy} to evaluate the asymptotic part
of $P(n)$.


For our following investigations, we rewrite \eq{Zprod} in the form
\beq
Z(\beta) = \exp \left\{\sum_{m=1}^{M}\ln\left[1+e^{-m^2\beta}\right]\right\}.
\label{Z2}
\eeq
In principle $M$ is infinity according to Eq.\ \eq{Zprod}. However, when calculating 
the partition density $g(E)$ using the Laplace inverse \eq{LapP} at finite $E$, 
we have for the reason given after Eq.\ \eq{partsum} the restriction
\beq
M(E)=[\sqrt{E}\,]\,,
\label{constr}
\eeq
where $[\sqrt{E}\,]$ denotes the largest integer $M$ contained in $\sqrt{E}$.

We note that $Z(\beta)$ in \eq{Z2} has no poles on the right half $\beta$ plane, 
including the imaginary axis as long as $M(E)$ is finite. We may therefore 
shift the contour $C$ onto the imaginary axis (i.e., choose $\epsilon=0$) and 
write the inverse Laplace transform \eq{LapiP} as
\beq
g(E) = \frac{1}{2\pi} \int_{-\infty}^{+\infty} 
       Z(i\tau)\,e^{iE\tau}\,{\rm d}\tau 
     = \frac{1}{2\pi} \int_{-\infty}^{+\infty} 
       \exp\left\{iE\tau+\ln Z(i\tau)\right\}{\rm d}\tau\,.
\label{geLapi}
\eeq
Since the imaginary part of the integrand above is antisymmetric w.r.t.\ $\tau=0$, 
the result $g(E)$ becomes real, as it must be, and we only need retain the real 
part of the integral:
\beq
g(E) = \frac{1}{2\pi} \int_{-\infty}^{+\infty} {\cal R}e\,  
       \exp\left\{iE\tau+\ln Z(i\tau)\right\}{\rm d}\tau\,.
\label{Pnint}
\eeq

We next define the Fourier transform (FT) of $g(E)$ by
\beq
{\cal F}_\tau [g(E)] = \int_{-\infty}^{+\infty} g(E)\,e^{-iE\tau}\,{\rm d}E
                     = F(\tau)\,.
\label{FourP}
\eeq
Note that $E$ and $\tau$ are a pair of conjugate dimensionless variables. 
Comparing \eq{LapP} and \eq{FourP}, we can write $F(\tau)$ as
\beq
F(\tau) = Z(i\tau)\,,
\label{FourPlim}
\eeq   
i.e., $F(\tau)$ is given by the values of $Z(\beta)$ along the imaginary axis 
$\tau$. This is a general property valid for any partition function $Z(\beta)$ which 
has no poles on or to the right of the imaginary axis. The absolute value of 
$F(\tau)$, inserting the partition function \eq{Z2} into \eq{FourPlim}, becomes
\beq
|F(\tau)| = \exp\left\{{\cal R}e\,\ln Z(i\tau)\right\}.
\label{FTabs}
\eeq
This Fourier spectrum will be studied numerically in Sec.\ \ref{secft}.


\subsection{Integral representation of $P(n)$}
\label{secatheo}

We start from the expression \eq{geLapi} for the Laplace inversion of $Z(\beta)$. 
We now formulate the following {\bf Lemma:} 
{\it The Laplace inversion integral \eq{geLapi} limited to the interval $\tau\in 
(-k\pi,+k\pi)$ with $k=1,2,\dots$ yields a sum of Bessel functions weighted by 
$P(n)$ such that its value at $E=n$ is $kP(n)$:}
\beq
g^{(k)}(E) = \frac{1}{2\pi} \int_{-k\pi}^{+k\pi} e^{iE\tau} Z(i\tau)\,{\rm d}\tau
       = k \sum_{n=0}^{\infty} P(n)\, j_0(k\pi(E-n))\,. \quad (k=1,2,3,\dots)
\label{lem}
\eeq
{\bf Proof:}
For finite $k$, we use the form \eq{Zgen} of $Z(\beta)$ and the integration
yields immediately Eq.\ \eq{lem}. Since $j_o(0)=1$, we get
\beq
g^{(k)}(E\!=\!n) = kP(n)\,,
\eeq
as claimed at the end of the lemma. {\it q.e.d.}\\
In the limit $k\to\infty$, the Bessel functions become delta functions
(see, e.g., Schiff \cite{schiff}):
\beq
\lim_{k\to\infty} [kj_0(k\pi(E-n))] = \delta(E-n)\,.
\eeq
Therefore the limit $k\to\infty$ yields the exact partition density as it should:
\beq
\lim_{k\to\infty} g^{(k)}(E) = \sum_{n=0}^\infty P(n)\, \delta(E-n) = g(E)\,.
\eeq

\begin{figure}[h]
\centering
\includegraphics[width=11cm,angle=0]{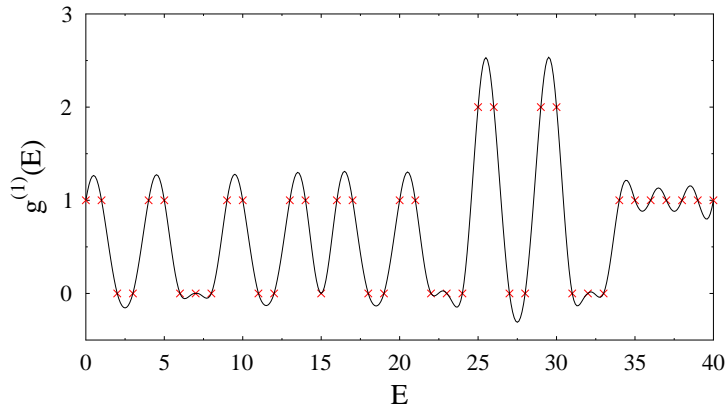}
\vspace*{-0.3cm}
\caption{Bessel-smoothed partition density $g^{(1)}(E)$ \eq{lem} (black line) for small
energies $E$. The red crosses at integer values of 
$E=n$ show the exact values of $P(n)$.} 
\label{figbes1}
\end{figure}

\begin{figure}[h]
\centering
\vspace*{-0.3cm}
\includegraphics[width=11cm,angle=0]{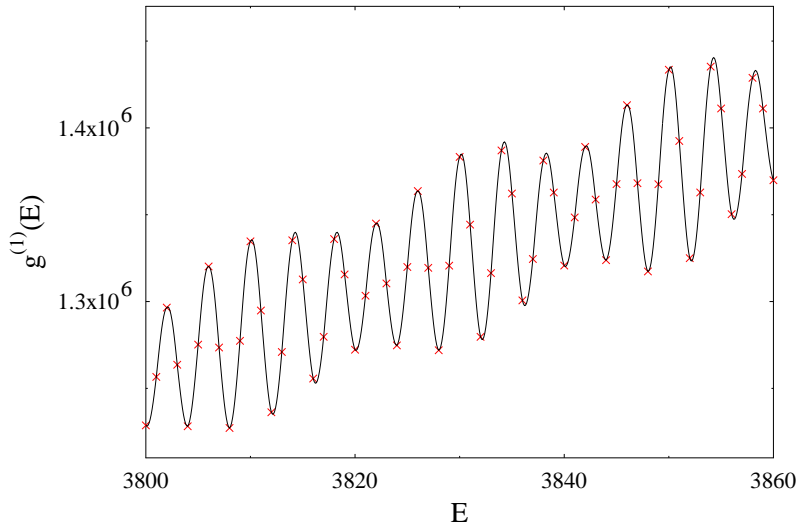}
\vspace*{-0.6cm}
\caption{The same as in Fig.\ \ref{figbes1} for larger energies $E$.}
\label{figbes2}
\end{figure}

The function $g^{(1)}(E)$ in \eq{lem} represents a smooth interpolation
curve through the exact values $P(n)$ at $E=n$, which we like to call a 
`Bessel-smoothed' partition density. It is shown in Figs.\ \ref{figbes1}
and \ref{figbes2}.

Setting $E=n$ in \eq{lem} with $k=1$, we obtain the following integral 
representation of $P(n)$:
\beq 
P(n) = \frac{1}{2\pi} \int_{-\pi}^{\pi} {\cal R}e\,
       e^{in\tau} Z(i\tau)\,{\rm d}\tau
     = \frac{1}{2\pi} \int_{-\pi}^{\pi} {\cal R}e\,
       \exp\left[in\tau+\sum_{m=1}^{[\sqrt{n}\,]}
       \ln\left(1+e^{-im^2\tau}\right)\right]\!{\rm d}\tau.
\label{intrepP}
\eeq
This integral formula is one of the central results of our paper. It will be 
the starting point in Sec.\ \ref{ssectrf} for the derivation of a semiclassical 
trace formula for the oscillations in $P(n)$.


\subsection{A continuous trace formula for $g(E)$}
\label{ssecctf}

We rewrite the real part of the exponent in (\ref{Pnint}):
\beq
g(E) = \frac{1}{2\pi} \int_{-\infty}^{+\infty} 
       \exp\left[{\cal R}e\,\ln Z(i\tau)\right]\,
       \cos[E\tau+{\cal I}m\,\ln Z(i\tau)]\,{\rm d}\tau\,.
\label{Pnint2}
\eeq
Using \eq{FTabs}, we obtain the following expression for the partition density:
\beq
g(E) = \frac{1}{2\pi} \int_{-\infty}^{+\infty} 
       |F(\tau)|\,\cos[E\tau+\phi(\tau)]\,{\rm d}\tau\,,
\label{PnintF}
\eeq
where the phase function $\phi(\tau)$ is given by
\beq
\phi(\tau) = {\cal I}m\,\ln Z(i\tau)
           = \sum_{m=1}^{[\sqrt{E}\,]} {\rm arctg} \left[
             \frac{-\sin(m^2\tau)}{1+\cos(m^2\tau)}\right].
\label{phi}
\eeq

In its structure, the expression \eq{PnintF} resembles a semiclassical trace formula 
(see \cite{book} for more details), with amplitudes given by the Fourier spectrum 
$|F(\tau)|$, actions by $E\tau$, periods by $\tau$, and phases by $\phi(\tau)$. The 
difference compared to the standard trace formulae \cite{gutz,book} is that here we do 
not have a discrete sum, but a continuous integral over periodic orbits labeled by the 
period variable $\tau$. We call Eq.\ \eq{PnintF} a `continuous trace formula' for the 
density $g(E)$ of F2 partitions. Note that it is exact, since no approximation has 
been made whatsoever in deriving it. It is, in fact, a general result for any level 
density $g(E)$, provided one knows its Fourier spectrum $|F(\tau)|$ and its Laplace 
transform on the imaginary axis $\tau$. Unless these ingredients are known analytically, 
Eq.\ \eq{PnintF} is of no practical use. The only exact case we found is the example of 
the one-dimensional harmonic oscillator without zero point energy. It has the 
single-particle spectrum $E_n=n$ $(n=0,1,2,\dots)$ and corresponds to $P(n)=1$. Its
Fourier peaks are at $\tau_k=2\pi k$ with $\phi(\tau_k)=0$, so that \eq{PnintF} 
reproduces the exact trace formula given in \cite{book}, eq.\ (3.40):
\beq
g_{ho}(E)=1+2\sum_{k=1}^\infty \cos(2\pi kE)\,.
\eeq

Eq. \eq{PnintF} gives us, however, a hint as to how a trace formula with discrete
orbits and periods could look like. In Sec.\ \ref{ssectrf} we shall use a semiclassical
approach for its realization. 


\section{Asymptotic expansion of the smooth part of $g(E)$}
\label{secasy}

In this section we briefly re-derive the leading asymptotic form $P_{as}(n)$ of the 
square partitions, given already in \cite{tran}, and then obtain a series of 
next-to-leading order contributions which yields a faster convergence to the 
average values of the exact $P(n)$. We closely follow the method used in 
Refs.\ \cite{tran,primep}. As stated above in 
\eq{averages}, averaging over $g(E)$ yields an average of $P(n)$. Therefore
the results of the smooth asymptotic part of $g(E)$ can be identified with
that of $P(n)$ when $E$ is taken as an integer $n$.

\subsection{Leading asymptotic form $g^{(0)}_{as}(E)$}
\label{ssecasy0}

We rewrite the inverse Laplace transform in Eq.\ \eq{LapiP} for the function
$g(E)$ in the form
\beq
g(E) = {\cal L}^{-1}_E [Z(\beta)] 
     = \frac{1}{2\pi i}\int_C e^{S(E,\beta)}\,{\rm d}\beta\,,
\label{LapiP2}
\eeq
where $S(E,\beta)$ is given by
\beq
S(E,\beta) = E\beta + \ln Z(\beta)\,.
\label{Sinf}
\eeq
which is often called the `entropy'. We now derive the asymptotic smooth part of 
$g(E)$ by performing the inverse Laplace transform \eq{LapiP2} by saddle-point 
integration. The asymptotic 
smooth part for large $E$ comes from the neighborhood of a real saddle point in the 
complex $\beta$ plane, lying near the imaginary axis. Doing the Laplace integral 
over such a saddle point by the method of stationary phase should yield the 
asymptotic smooth part of $g(E)$. Hereby we can approximate the sum over $m$
in \eq{Z2} using the Euler-MacLaurin expansion \cite{abro}, yielding
\beq
S(E,\beta) \simeq E\beta + \int_{0}^{\infty}dq\ln\left[1+e^{-q^2\beta}\right]
                         - \frac{1}{2}\ln(2)\,.
\eeq
Doing the integration \cite{gr}, we obtain for real $\beta=x$
\beq
S(E,x) \simeq Ex +\frac{\Gamma(3/2)\eta(3/2)}{\sqrt{x}}
           -\frac{1}{2}\ln(2)\,, \qquad
           \eta(3/2)=\sum_{l=1}^\infty \frac{(-1)^{l-1}}{l^{3/2}}\,,
\eeq
where $\eta(z)$ is the Dirichlet eta function.
To find a real saddle point (SP) $x_0$, we have to solve the SP equation
\beq
S_1(E,x_0) = \left.\frac{\partial S(E,x)}{\partial x}\right|_{x_0}
               = E-\frac{D}{2{x_0}^{3/2}}=0\,, \quad
                 D = \Gamma(3/2)\eta(3/2)=0.678093895.
\eeq
Here $S_i(E,x)$ denotes the $i$-th partial derivative of $S(E,x)$ with respect to 
$x$ at fixed $E$. This yields the solution for the SP $x_0$ at each energy $E$:
\beq
x_0(E) = \left[\frac{D}{2E}\right]^{2/3} 
       = \lambda_0 E^{-2/3}\,, \qquad \lambda_0 = (D/2)^{2/3} =0.486227919\,.
\label{x0}
\eeq
Doing the contour integral over $\tau$ parallel to the imaginary axis in the
stationary-phase approximation yields
\beq
g^{(0)}_{as}(E) = \frac{\exp[S(E,x_0)]} {\sqrt{2\pi |S_2(E,x_0)|}}
 = \sqrt{\frac{\lambda_0}{6\pi}}\,E^{-5/6}\,\exp\left(3\lambda_0 E^{1/3}\right),
\label{g0as}
\eeq
so that, using the property \eq{averages},
\beq 
P^{(0)}_{as}(n) = \sqrt{\frac{\lambda_0}{6\pi}}\,n^{-5/6}\,
                  \exp\left(3\lambda_0 n^{1/3}\right).
\label{pas0}
\eeq
This is the leading-order asymptotic form of $P(n)$ given already in Ref.\ 
\cite{tran} (where our present $\lambda_0$ was called $\lambda_2$).

\subsection{Higher-order contributions to $g^{(0)}_{as}(E)$}
\label{ssecspcor}

\begin{figure}[h]
\centering
\vspace*{-0.3cm}
\includegraphics[width=12.cm,angle=0]{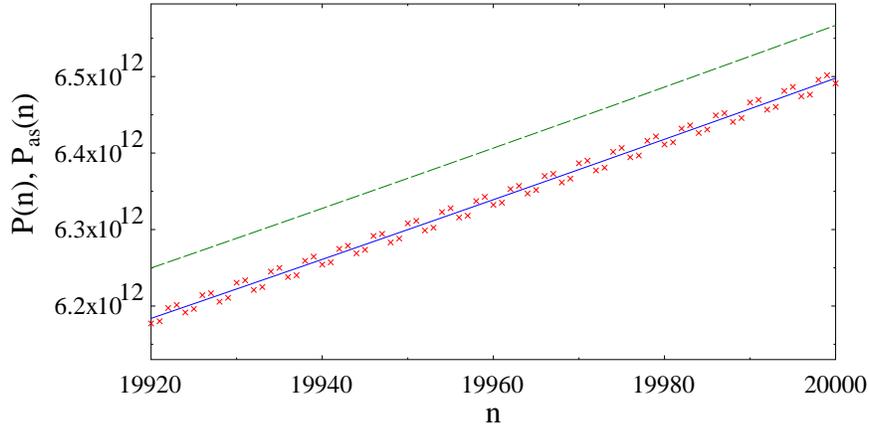}
\vspace*{-0.7cm}
\caption{Exact $P(n)$ by crosses (red), leading-order asymptotic part 
$P^{(0)}_{as}(n)$ \eq{pas0} by the dashed (green) line, and corrected 
asymptotic part $P_{as}(n)$ \eq{pas1} by the solid (blue) line in the 
large-$n$ region.}
\label{figpas}
\end{figure}

In Appendix \ref{secspcor} we derive higher-order saddle-point contributions to 
the result \eq{g0as}, yielding a better asymptotic form $g_{as}(E)$ up to third
order as given in \eq{apx22}. For the F2 partition counting function, this yields
\beq
P_{as}(n) = \sqrt{\frac{\lambda_0}{6\pi}}\,n^{-5/6}\,e^{3\lambda_0 n^{1/3}}
            \left[1-c_1\, n^{-1/3}-c_2\, n^{-2/3}-c_3\, n^{-1}\right],
\label{pas1}
\eeq
with the coefficients $c_1$, $c_2$, and $c_3$ given in Eqs.\ (\ref{c12}), (\ref{c3}).
In Fig.\ \ref{figpas} we show the leading approximation $P^{(0)}_{as}(n)$ (dashed
line) and the improved result $P_{as}(n)$ using \eq{pas1} (solid line). The 
latter is seen to give an excellent agreement with the average through the 
exact $P(n)$.

\begin{figure}[h]
\centering
\vspace*{-0.2cm}
\includegraphics[width=12.cm,angle=0]{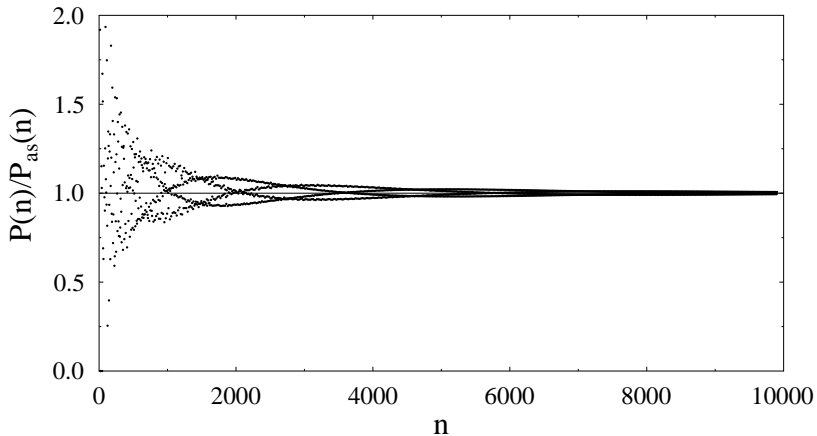}
\vspace*{-0.6cm}
\caption{Ratio $P(n)/P_{as}(n)$ showing that $P_{as}(n)$ in \eq{pas1} including the 
SP corrections reaches correctly the average values of $P(n)$ in the large-$n$ 
limit: $P(n)/P_{as}(n)\rightarrow 1$.
}
\label{figpasrat}
\end{figure}

Figure \ref{figpasrat} shows that the relative amplitude of the oscillations 
decreases exponentially with $n$, as given more quantitatively in Eq.\ \eq{dPbyPas} 
below, and that the ratio $P(n)/P_{as}(n)$ reaches unity at $n\simg 10,000$. 
The improved asymptotic form \eq{pas1} thus brings a substantial improvement 
over the leading-order term $P^{(0)}_{as}(n)$ for which the limit 
$P(n)/P_{as}(n)\rightarrow 1$ is not reached (see also Ref.\ \cite{oeis}).

For $n\simg 100,000$ we found that the truncated series with three terms in 
\eq{pas1} does not converge fast enough; we will come back to this point at the 
end of Sec.\ \ref{ssectrf}.


\section{Fourier analysis of $g(E)$} 
\label{secft}

In order to understand the oscillating part of $P(n)$ we now investigate the 
Fourier spectrum Eq.\ \eq{FTabs} of the F2 partitions. We shall use both forms 
given in Eqs.\ \eq{Zgen} and \eq{Z2} for $Z(\beta)$ and plot the results versus 
the frequency $f=2\pi/\tau$.

\subsection{Fourier spectra using Eq.\ \eq{Zgen}}
\label{ssecftZgen}

We first look at the Fourier spectra obtained from \eq{Zgen} with a linear cut-off
given by
\beq
F(f) = \sum_{n=0}^{n_{max}} P(n)\,e^{-2\pi i\,n/f}\,(1-n/n_{max})\, .
\label{FTZgen}
\eeq
\begin{figure}[h]
\centering
\vspace*{-0.9cm}
\includegraphics[width=11.5cm,angle=0]{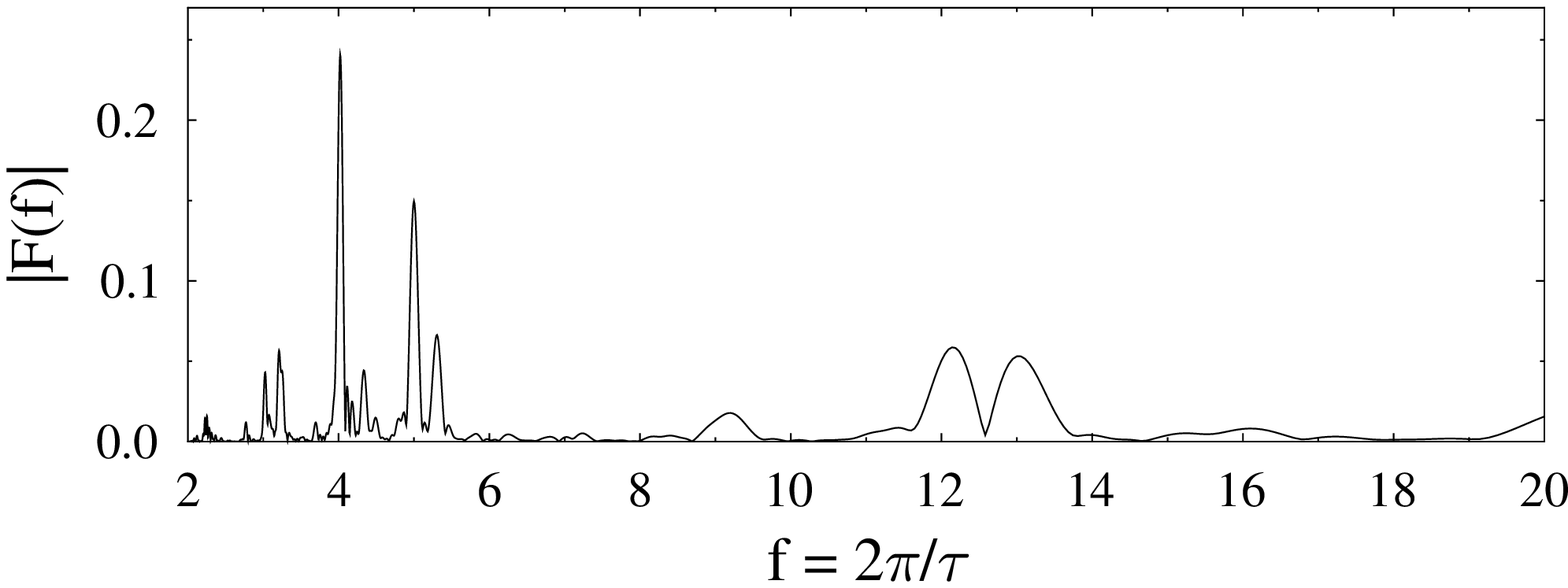}\vspace*{-0.5cm}
\includegraphics[width=11.5cm,angle=0]{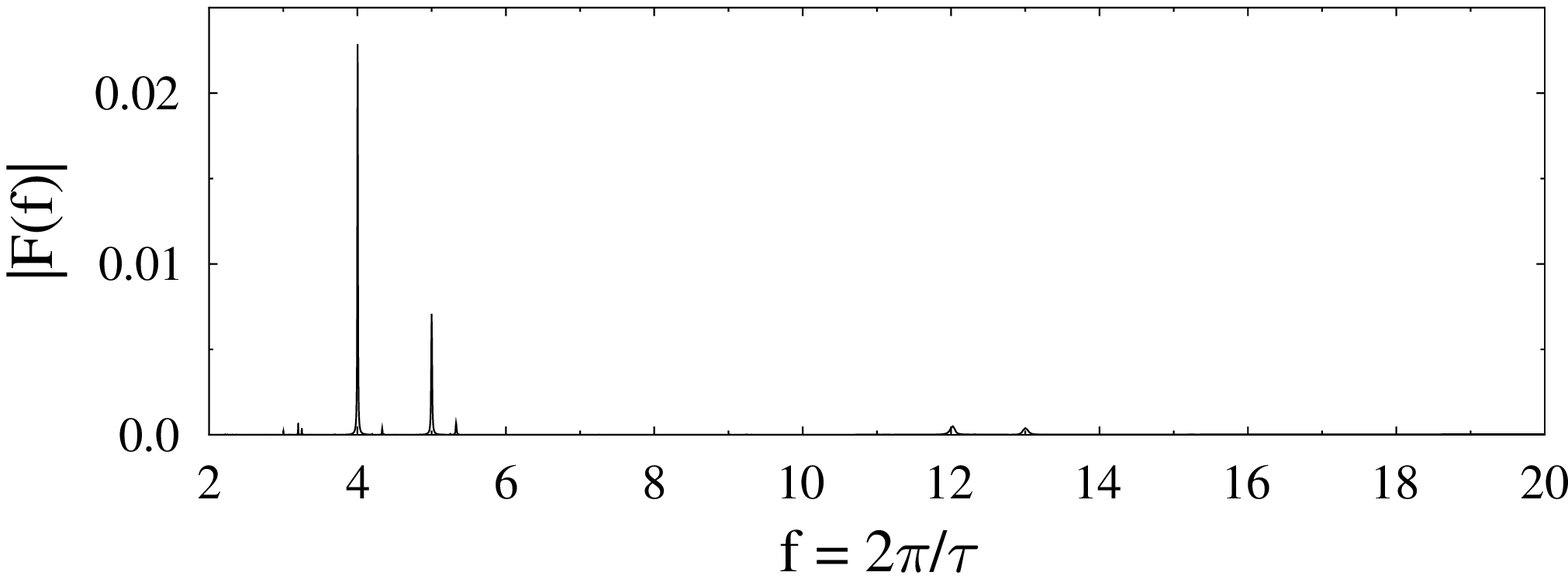}\vspace*{-0.5cm}
\includegraphics[width=11.5cm,angle=0]{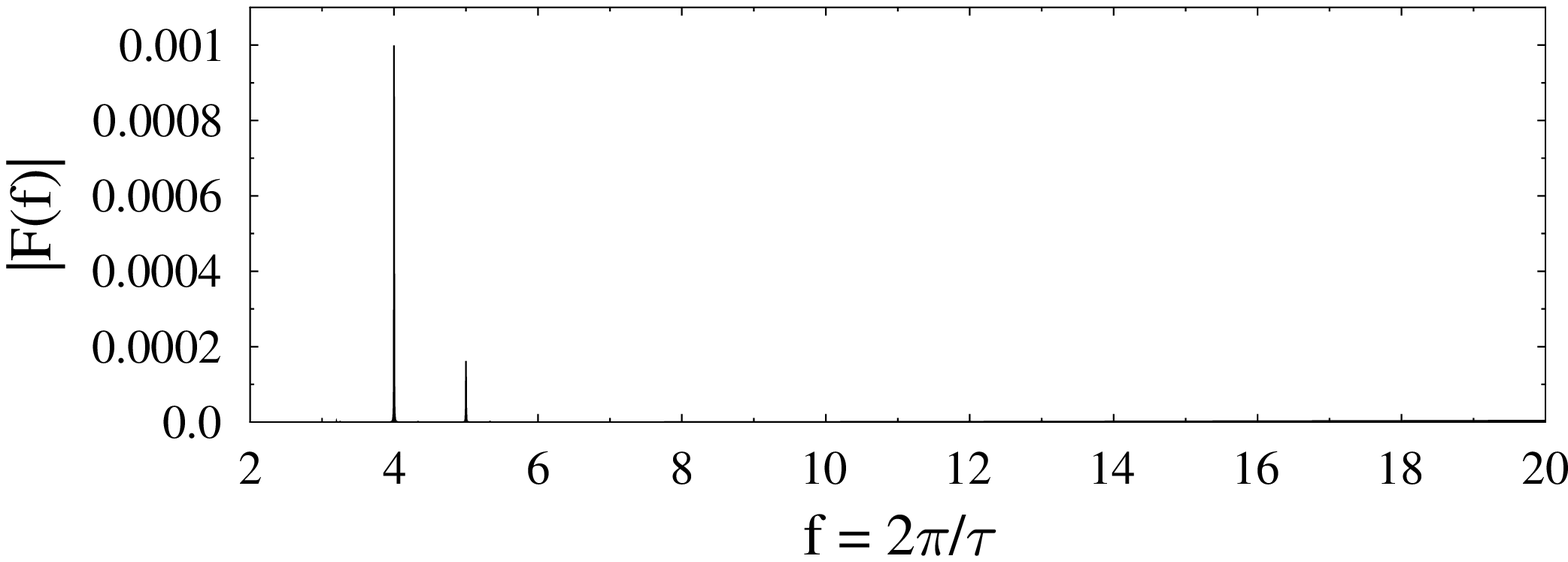}\vspace*{-0.5cm}
\caption{Fourier spectra $|F(f)|$ using $Z(\beta)$ in \eq{Zgen} versus frequency $f$. 
The cut-off of the $n$ sum in \eq{FTZgen} is, from top to bottom, $n_{max}$ = 300, 4000, 
and 20,000.
}
\label{figFTP}
\end{figure} 

Figure \ref{figFTP} shows the results for $|F(f)|$ with, from top to bottom, $n_{max}$ 
= 300, 4000, and 20,000 in the region $2 \leq f \leq 20$ (normalized to unit intensity
at $f=1$). The spectra are clearly cut-off dependent and reveal a varying number of 
visible frequencies in the different regions of $n$. For the lowest cut-off, the peaks 
are becoming more 
diffuse with increasing $f$, but we clearly find many more frequencies than for the 
higher cut-offs. For $n_{max}=300$ we recognize peaks, with varying intensities, near
the frequencies $f$ = 4, 5, 12, and 13 (besides others). Interestingly, these numbers 
belong to the lowest PTs (3,4,5) and (5,12,13). This gives us a hint that the 
spectrum of $P(n)$ is dominated by the lowest PTs. For $n_{max}=4000$, the peaks 
at $f$ = 4 and 5 dominate, while only tiny hints of the other frequencies can be seen. 
For $n_{max}=20,000$ no more trace is left of the other frequencies, and the frequency 
4 is clearly dominating over $f=5$. 

The tendency that a decreasing number of frequencies is important with increasing
$n$ will be confirmed in Sec.\ \ref{sectrf} by the semiclassical interpretation of 
the oscillations in $P(n)$.

\subsection{Fourier spectra using Eq.\ \eq{Z2}}
\label{ssecftZ2}

Because of the bad resolution of the above spectra for small $n$, we now 
investigate the FT spectra that we obtain using Eq.\ \eq{Z2} for $Z(\beta)$. 
Their frequency information is in principle the same, but the peaks turn out to 
be much cleaner, allowing us to identify their exact frequencies $f$ and to compare 
their relative intensities.

We first discuss the dependence on the upper limit $M$ of the $m$ summation in \eq{Z2}. 
We argue that $|F(f)|$ actually is a distribution function with infinite peaks at many 
integer or rational values of $f$. It therefore has to be normalized in some suitable 
way. 

\subsubsection{Normalization of $|F(f)|$}
\label{ssecftnorm}

\begin{figure}[h]
\centering
\vspace*{-0.2cm}
\includegraphics[width=8cm,angle=0]{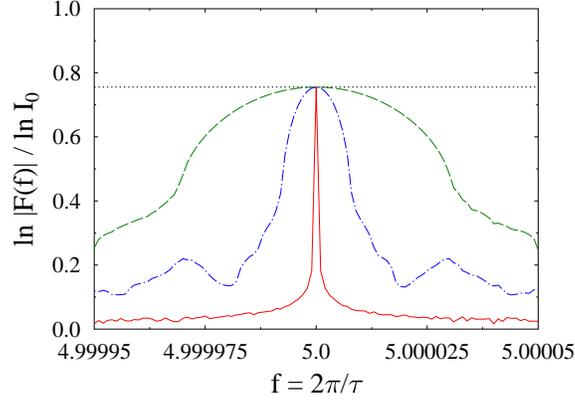}\vspace*{-0.5cm}
\caption{Close-up of the scaled $\ln|F(f)|$ near the frequency $f=2\pi/\tau=5$, 
shown for $M=$ 500 by the dashed (green) line, $M=$ 1000 by the dash-dotted 
(blue) line, and $M=$ 5000 by the solid (red) line. The horizontal dotted line 
gives the theoretical peak height according to Tab.\ \ref{tabgen}.}
\label{figFT5}
\end{figure}

In Fig.\ \ref{figFT5} we show the intensity of $|F(f)|$ in a close-up near a 
peak at $f=5$. Shown are the curves for increasing values of the upper limit 
$M$ of the $m$ summation, namely for $M=500$, 1000, 
and 5000 (from top to bottom). The horizontal line gives the calculated value 
0.7554 of the scaled intensities (see Tab.\ \ref{tabgen} below). We see that with 
increasing $M$, the curves become narrower, and in the limit $M\to\infty$, we can 
take the local spectrum to be a (renormalized) delta function peaked at $f=5$.

The normalization of the curves in all figures has been chosen in the 
following way. It is easy to see that for $\tau_k=2k\pi$ with integer $k$, the 
exponential in Eq.\ \eq{Z2} becomes unity, so that we obtain $|F(\tau_k)|=
\exp(M\ln 2)$. Below we will call the peaks at $\tau_k=2k\pi$ (i.e.\ $f_k=1/k$) 
the `generation 0' peaks which on a logarithmic scale have the intensity 
$\ln I_0=M\ln 2$. In all figures we therefore show the scaled function 
$\ln|F(f)|/\ln I_0$, so that the generation 0 peaks have the height unity
(even if not seen for $f\geq 2$).

\subsubsection{Numerical Fourier spectra $|F(f)|$}
\label{ssecftnum}

\begin{figure}[h]
\centering
\vspace*{-1.25cm}
\includegraphics[width=12.5cm,angle=0]{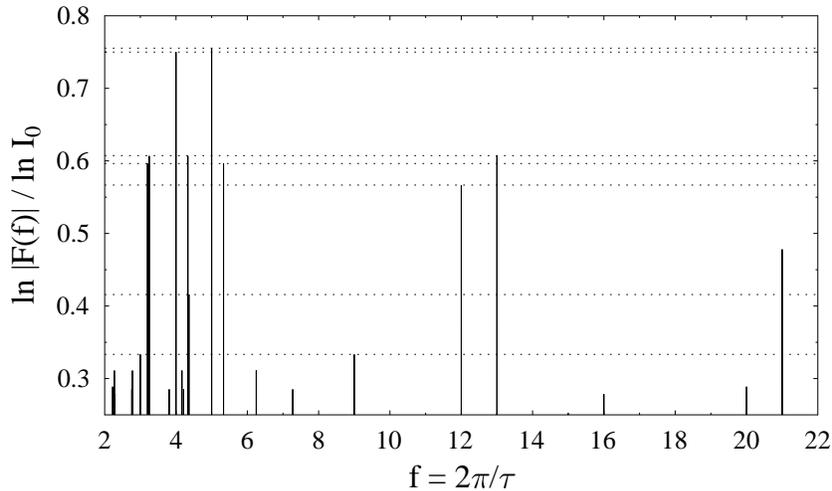}\vspace*{-1.1cm}
\caption{Scaled Fourier transform $\ln|F(f)|$ of $g(E)$ on a logarithmic vertical
scale. The horizontal dashed lines give the calculated relative intensities of 
the first 10 generations.}
\label{figftr}
\end{figure} 

Figure \ref{figftr} shows the Fourier spectrum as a function of the frequency 
$f=2\pi/\tau$, plotted for $2\leq f \leq 22$ with the normalization defined above. 
The peaks are very sharp. We find peaks located exactly at $f=3$, 4, 5, 9, 12, 13, 16,
20, and 21; all other peaks in this interval appear at rational frequencies. We can now
classify the peaks into `generations' with decreasing intensities, as listed in Tab.\ 
\ref{tabgen} below. The vertical scale of Fig.\ \ref{figftr} was selected such that 
the peaks of the generations 1 - 10 can be clearly differentiated; their theoretical 
scaled intensities as given in Tab.\ \ref{tabgen} are shown by the horizontal dashed 
lines. 

\begin{figure}[h]
\centering
\vspace*{-0.8cm}
\includegraphics[width=13cm,angle=0]{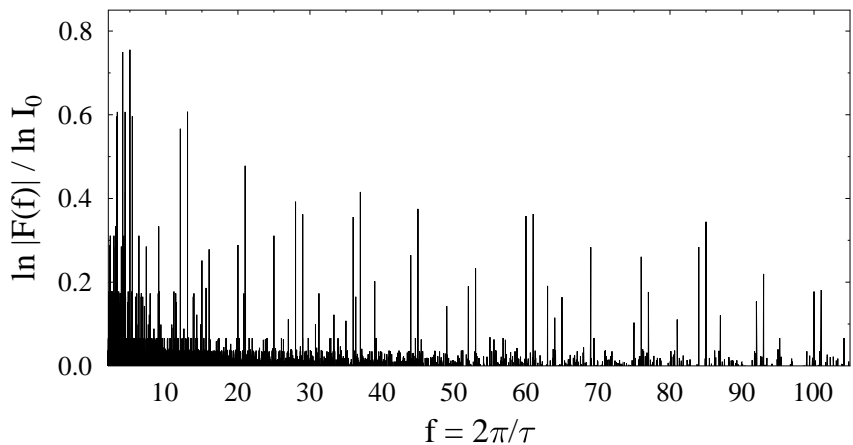}
\vspace*{-0.8cm}
\caption{Same as Fig.\ \ref{figftr} over a larger range of the frequency $f$. 
}
\label{figfthi}
\end{figure}

Figure \ref{figfthi} shows the same for $f$ up to 105. Many more integer-valued 
frequencies appear. We notice, in particular, the dominating
intensities of peak pairs with the frequencies (4,5), (12,13), (28,29), (36,37), 
(60,61), (84,85), and (100,101). Four of them appear as the largest numbers in 
PTs, namely in (3,4,5), (5,12,13), (11,60,61) and (13,84,85). The numbers 28, 29
and 101 appear isolated in other PTs. 

We have thus found a strong evidence that the PTs play a dominant role in the 
spectrum and hence also the oscillations of $P(n)$. 
This will, indeed, be confirmed quantitatively by the
semiclassical trace formula derived and discussed in Sec.\ \ref{sectrf}.


For reasons that become evident in Sec.\ \ref{sectrf}, we shall now map the
interval $\tau\in(0,2\pi)$ on the interval $\tau\in(-\pi,+\pi)$. The Fourier
spectrum then is symmetric about $\tau=0$, with the peak pairs of each generation
appearing with opposite signs of $\tau$. 

Table \ref{tabgen} presents the first ten generations of Fourier peaks and their 
properties; for the periods $\tau_g$ only the positive values are given. Generation 
0 creates the smooth part $P_{as}^{(0)}$, as discussed in Sec.\ \ref{ssecasy0}. For 
convenience and brevity, the constants $\lambda_g$, $\mu_g$, $\kappa_g$, and 
$\varphi_g$ are given in the table but will be discussed later in Sec.\ 
\ref{ssecsp} together with the use of the generations 1-10 as `orbits' in a 
semiclassical trace formula.

\Table{tab2}{13.5}{
\begin{tabular}{|c|c|c|c|c|c|c|c|}\hline
\;g\; &$\tau_g$ & $f_g$ & \,$\ln I_g/\!\ln I_0\,$ & \,$\lambda_{g}$ 
  &$\mu_g$ & \,$\kappa_{g}$ & $\varphi_g$ \\ 
\hline
0 &0         & 1$^\dagger$& 1.0    &\,0.48622\, &\,0          &\,3.085\,  &  0       \\
1 &$2\pi/5$  & 5        & 0.7554 &\,0.37444\, &\,0.0000\,   &\,4.006\,  &  0.0000  \\
2 &$2\pi/4$  & 4        & 3/4    &  0.400     &\,0.1999187\,&\,3.352\,  &\,-0.2318 \\
3 &$2\pi/13$ & 13       & 0.6072 &  0.30743   &\,0.0000\,   &\,4.877\,  &  0.0000  \\
4 &$6\pi/13$ & 13/3     & 0.6072 &  0.30743   &\,0.0000\,   &\,4.877\,  &  0.0000  \\ 
5 &$8\pi/13$ & 13/4     & 0.6072 &  0.30743   &\,0.0000\,   &\,4.877\,  &  0.0000  \\
6 &$6\pi/16$ & 16/3     & 0.5964 &  0.3200    &\,-0.154191\,&\,4.219\,  &  0.2242  \\
7 &$10\pi/16$& 16/5     & 0.5964 &  0.3200    &\,0.154191\, &\,4.219\,  &\,-0.2242 \\
8 &$2\pi/12$ & 12       & 0.5667 &  0.3129    &\,0.1606108\,&\,4.264\,  &\,-0.2358 \\
9 &$22\pi/48$& 48/11    & 0.4158 &  0.245     &\,-0.14152\, &\,6.502\,  &   0.2542 \\
10&$2\pi/3$  & 3        & 0.3333 &  0.2847    &\,0.364242\, &\,3.244\,  &\,-0.4538 \\
\hline
\end{tabular}}{~~Successive generations $g$ of Fourier peaks, their periods $\tau_g$, 
frequencies $f_g$ and scaled relative intensities $I_g$. The constants $\lambda_g$, 
$\mu_g$, $\kappa_g$, and $\varphi_g$ are defined and discussed in Sec.\ \ref{sectrf}.
($^\dagger$The frequency $f_0=1$ corresponds to $\tau=2\pi$ which also belongs to 
generation 0.)
\label{tabgen}
}

We should note that there exist further generations (with frequencies 9, 9/2, and 9/4) 
that have the same intensities as generation 10. Their contributions to the trace 
formula \eq{dgsc} are, however, negligible. We have listed the generation 10 here mainly 
because its frequency 3 belongs to the PT (3,4,5).

We also point out that the ordering of the generations is done here according
to the decreasing intensities $\ln I_g/\!\ln I_0$. In Sec.\ \ref{ssectrf} we shall
see that the semiclassical amplitudes $A_g(n)$ of the `orbits' appearing in the 
trace formula \eq{dgsc} follow a somewhat different ordering, in agreement with
the results in Fig.\ \ref{figFTP} where the peaks with $f=4$ (generation 2) have 
a higher intensity than those with $f=5$ (generation 1).

\newpage
 
\section{Derivation of a trace formula for $\delta P(n)$}
\label{sectrf}

In this section we derive a semiclassical trace formula for the 
oscillating part of $P(n)$. The main idea of our approach is that 
asymptotic expressions of oscillating functions can be found from 
stationary-phase integration over complex saddles in the $\beta$ plane. 
As shown in Sec.\ \ref{ssecasy0}, the asymptotic smooth part $P_{as}(n)$ 
is obtained from the real saddle point $x_0$ given in \eq{x0}. This 
technique has been used, e.g., by Balazs {\it et al.}\ \cite{BPD} to 
find both smooth and oscillating asymptotic expressions for integrals of 
the Airy function. We shall illustrate this method in Appendix \ref{secairy} 
for the Airy function itself.

We start from the integral representation \eq{intrepP} of $P(n)$ for which the 
integration along the $\tau$ axis from $-\pi$ to $+\pi$ (contour $C$) yields
the exact $P(n)$ for all $n$. Since the integrand has no singularities for $x>0$, we 
may deform the contour arbitrarily, keeping its end points fixed. We choose 
it to pass over the most important saddles in the complex $\beta$ plane 
corresponding to the leading Fourier peaks, and then use stationary-phase 
integration locally at each saddle. 
\begin{figure}[h]
\centering
\vspace*{-0.3cm}
\includegraphics[width=5.5cm,angle=0]{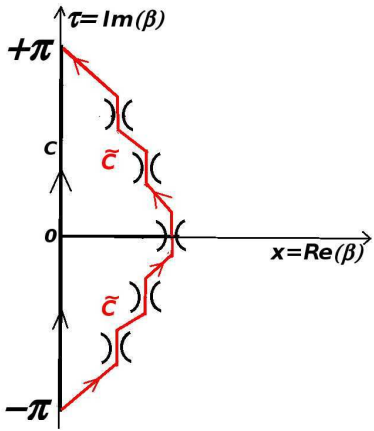}
\vspace*{-0.3cm}
\caption{Schematic plot of the contour for the integral \eq{intrepP}. The exact 
contour $C$ (going from -$\pi$ to +$\pi$ along the $\tau$ axis) is deformed into 
a new contour (red) ${\widetilde C}$ which goes over some selected saddles in the 
complex $\beta$ plane.}
\label{figschem}
\end{figure}

Figure \ref{figschem} shows a sketch of the situation in the complex $\beta$ plane, 
with the deformed contour ${\widetilde C}$ chosen to pass over 5 representative 
saddles. The exact path between the saddles does not matter, since we will only 
collect the local contributions near each saddle in the stationary-phase 
approximation. While the real saddle yields the smooth $P_{as}(n)$ as shown in 
Sec.\ \ref{ssecasy0}, the complex saddles will yield an approximation for the 
oscillating part $\delta P(n)$.

\subsection{Saddle points in the complex $\beta$ plane}
\label{ssecland}

Scanning the complex $\beta$ plane for the function ${\cal R}e\,S(E,\beta)$ in 
\eq{Sinf} using Eq.\ \eq{Z2}, we observe that for each of the dominant Fourier
peaks listed in Tab.\ \ref{tabgen}, there exists a saddle at a stationary
point with $\tau$ close to the value $\tau_g$ listed there. We illustrate this
in the following three figures for the generations 0, 1, and 2, with contour 
plots of the function ${\cal R}e\,S(E,\beta)$ (evaluated here at $E=40$) in the 
complex $\beta$ plane. 

\begin{figure}[h]
\centering
\hspace*{-0.5cm}
\includegraphics[width=6.6cm,angle=-90]{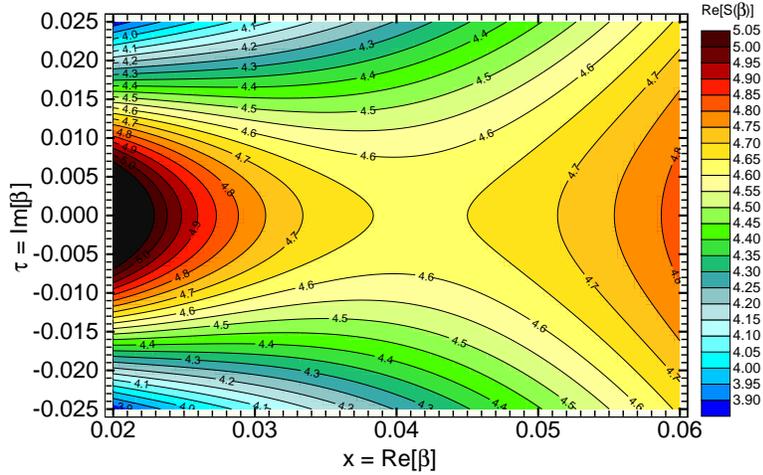}
\caption{Contour plot of the function ${\cal R}e\,S(E,\beta)$ in \eq{Sinf}, evaluated 
at $E=40$, in the saddle-point region of generation 0 at $\tau=0$.}
\label{figsad0}
\vspace*{-0.2cm}
\end{figure}

Figure \ \ref{figsad0} shows the real saddle at $\tau_0=0$ which was used in Sec.\ 
\ref{ssecasy0} to derive the asymptotic function $P_{as}(n)$; the position of the 
saddle point $x_0$ axis is given in \eq{x0}. 
\medskip

\begin{figure}[h]
\centering
\hspace*{-0.5cm}
\includegraphics[width=6.6cm,angle=-90]{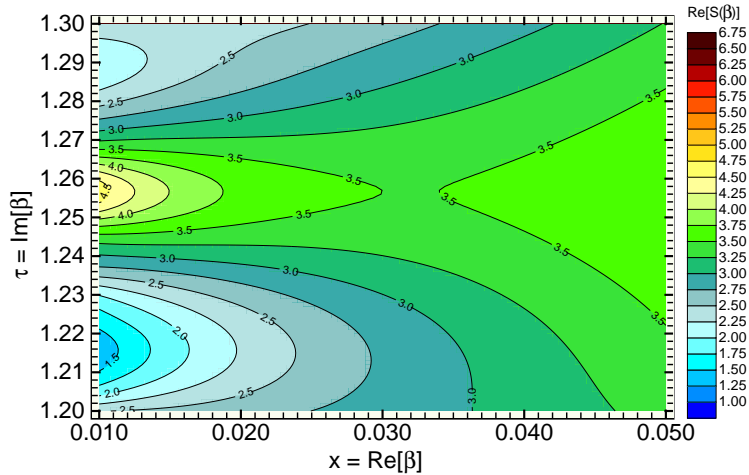}
\caption{Same as Fig.\ \ref{figsad0} for generation 1 at $\tau_1=2\pi/5$.}
\label{figsad1}
\end{figure} 

Figure \ref{figsad1} shows one of the saddles of generation 1, lying approximately 
at $\tau_1=2\pi/5$. Note that for generations 0 and 1, the path of steepest 
descent appears to be parallel to the imaginary axis. This is, however, analytically 
true only for the real saddle, as will be discussed in the following subsection.

\begin{figure}[h]
\centering
\vspace*{0.2cm}
\hspace*{-0.5cm}
\includegraphics[width=6.6cm,angle=-90]{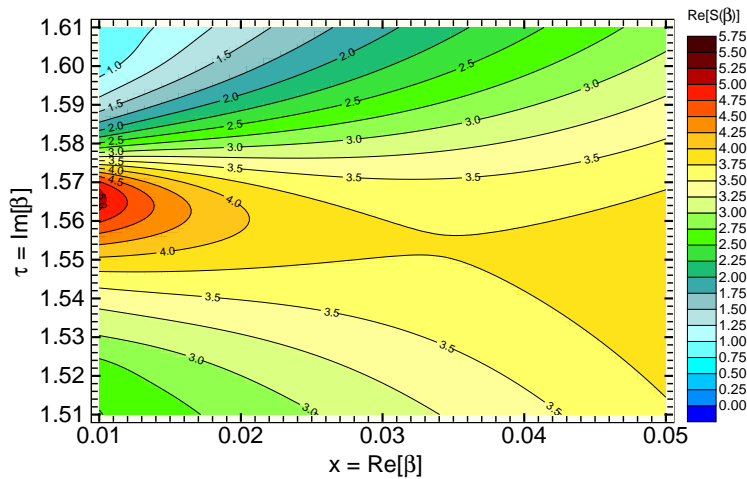}
\caption{Same as Fig.\ \ref{figsad0} for generation 2 at $\tau_2 \sim 2\pi/4$. 
Note that, different from the two previous cases, the saddle here is tilted
with respect to the $x$ and $\tau$ axes. }
\label{figsad2}
\end{figure}

Figure \ref{figsad2} shows one of the saddles of generation 2, lying somewhat below
$\tau_2=2\pi/4$. The precise value of $\tau_2$ at the saddle is energy dependent and
given in Eq.\ \eq{mug} below. The path of steepest descent here is tilted with 
respect to the $\tau$ axis by an angle $\alpha_2$ which leads to the Maslov index 
$\varphi_2$ defined in Eq.\ \eq{maslov} below.
 
What we see in these three examples can be summarized as follows. From each
of the dominant Fourier peaks listed in Tab.\ \ref{tabgen} (and for many more),
a ridge with a conditional maximum of $S(E,\beta)$ in the $\tau$ direction 
descends from its position on the $\tau$ axis (where it corresponds to a peak 
that diverges for $M\to\infty$ as discussed in Sec.\ \ref{ssecftnorm}) towards 
the right, i.e., in the $x$ direction. Eventually it will deviate from the $x$ 
direction (except for the generations 0, 1, and 3-5) and form a saddle at some 
value $x_g(E)$. The exact position $\beta_g(E)=x_g(E)+i\tau_g(E)$ of the saddle 
point (SP) of each generation (as a function of energy $E$) has to be found by 
solving the SP equations discussed next. 

\subsection{Solutions of the complex saddle-point equations}
\label{ssecsp}

When integrating over a saddle by the method of steepest descent, one chooses the 
direction in which the real part of the entropy function $S(E,\beta)$ in \eq{LapiP2}
has its steepest maximum. This yields a rapidly oscillating phase which tends to 
cancel all contributions except close to the SP where the phase becomes stationary 
(hence the alternative name `stationary-phase approximation'). Near the SP $\beta_0$, 
the integrand may be approximated by a Gaussian integral that can be evaluated exactly 
if the curvature of ${\cal R}e\,S(E,\beta)$ at $\beta_0$ is known. While this integral 
yields a smooth semiclassical amplitude as a function of $E$ (as in Sec.\ 
\ref{ssecasy0}), the imaginary part ${\cal I}m\,S(E,\beta)$ at $\beta_0$ yields an 
oscillatory function of $E$ if $\tau_0$ is not zero. 

The saddles are thus determined by the stationary condition $\left.{\partial 
S(E,\beta)}/{\partial \beta} \right|_{\beta_0}=0$ in the complex $\beta$ plane, 
which yields two equations. Using \eq{Sinf} and \eq{Z2} we find from the real part
\beq
E - \frac12 \sum_m m^2 \frac{ \left[\exp(-m^2x_0)+\cos(m^2\tau_0)\right] }
              {\left[{\rm Cosh}(m^2x_0)+\cos(m^2\tau_0)\right]}
  = \left.\frac{\partial}{\partial x}\,{\cal R}e\,S(E,\beta)\right|_{\beta_0}
  = 0\,, 
\label{resp}
\eeq
and from the imaginary part 
\beq
 -\frac12 \sum_m m^2 \frac{\sin(m^2\tau_0)}
          {\left[ {\rm Cosh}(m^2x_0)+\cos(m^2\tau_0)\right]} 
 = \left.\frac{\partial}{\partial\tau}\,{\cal R}e\,S(E,\beta)\right|_{\beta_0}
 = 0\,. 
\label{imsp}
\eeq
These equations may be transformed into a more symmetrical form, involving 
double sums
\begin{eqnarray}
E&=&\sum_m m^2\sum_{k=1}^\infty (-1)^{k+1}
e^{-km^2x_0}\cos(km^2\tau_0)\,,\\
0&=&\sum_m m^2\sum_{k=0}^\infty (-1)^ke^{-km^2x_0}\sin(km^2\tau_0)\,.
\label{imsp2} 
\end{eqnarray}
We were not able to solve these SP equations analytically, except for the case of the 
real saddle at $\tau_0=0$ for generation 0. In this case, \eq{imsp} and \eq{imsp2} are 
trivially fulfilled for all $x_0$ since $\sin(m^2\tau_0)=0$ for all $m$ (and $k$). For 
\eq{resp} we could then use the Euler-MacLaurin approximation (replacing the $m$ 
sum by an integral, see Sec.\ \ref{ssecasy0}) with the result given in \eq{x0}. 
In the same spirit, we found that our numerical solutions of the SP equations for 
all generations $g$ could be very accurately fitted by the equations
\bea
   x_0^{(g)}(E) & = & \lambda_g E^{-2/3}\,, \label{lamg}\\
\tau_0^{(g)}(E) & = & \pm\,\tau_g \mp \mu_g E^{-2/3}\,,
\label{mug}
\eea
with the values of $\lambda_g$ and $\mu_g$ given in Tab.\ \ref{tabgen}. For the
generation 0, $\mu_0$ is exactly zero. For generations 1 and 3-5, $\mu_g$ is zero
within the numerical accuracy of the constants given in the table.
The entropies $S$ at the saddle points could be fitted by
\bea
{\cal R}e\;S(E,\beta_0) & = & 3\lambda_g E^{1/3}-\ln(2)/2\,.\\
\label{ReSadg}
{\cal I}m\;S(E,\beta_0) & = & \pm E \tau_g \mp 3\mu_g E^{1/3}\,.
\label{ImSadg}
\eea
Note that the last term in \eq{ImSadg} containss a contribution
$\phi_g(E)$ defined by
\beq
\phi_g(E) = {\cal I}m\,\ln Z(\beta_g)
          = \sum_m {\rm arctg} \left\{
            \frac{-\exp[-m^2x_0^{(g)}(E)]\,\sin[m^2\tau_0^{(g)}(E)]}
            {1+\exp[-m^2x_0^{(g)}(E)]\,\cos[m^2\tau_0^{(g)}(E)]} \right\},
\label{phig}
\eeq
which is the analogue of the phase $\phi(\tau)$ in \eq{phi} but evaluated here 
at the complex saddles $\beta_g(E)$. It was numerically found to be well 
approximated by
\beq
\phi_g(E) = \mp\,2\mu_g E^{1/3}\,. 
\label{phigfit}
\eeq

For the determination of the direction of steepest descent, we proceed
as follows. From a given SP $(x_o,\tau_0$), we define a straight line in
the direction $\alpha$ by
\beq
\beta(r,\alpha) = (x_0+i\tau_0) + r\,e^{i\alpha}\,.
\label{ralpha}
\eeq
We then calculate for each $\alpha$ the curvature along $r$ by
\beq
K_r(E,\alpha) = \left.{\cal R}e\,\frac{\partial^2 S(E,\beta(r,\alpha))}
                {(\partial r)^2}\right|_{E,\alpha}\,.
\label{Kr}
\eeq
This becomes a periodic function whose minimum $\alpha_g\in(0,2\pi)$ for each 
generation yields the direction of steepest descent with a maximum absolute value 
of the (negative) curvature. The resulting $K_r$ could then be numerically 
fitted by the equation
\beq
K_r(E,\alpha_g) = -\kappa_g E^{5/3}\,,
\label{kapg}
\eeq
with the values $\kappa_g$ given in Tab.\ \ref{tabgen}. For a SP at $\tau_g$ with 
phase $\alpha_g$, the symmetry partner at $-\tau_g$ has the phase $\pi-\alpha_g$ due
to the antisymmetry of $S(E,\beta)$ with respect to the real axis.

All the above functions with the constants in Tab.\ \ref{tabgen}, representing the 
solutions of the SP equations \eq{resp} and \eq{imsp}, will henceforth only be used 
for integer values of the energy variable $E=n$. It remains a challenge for future
research to find analytical expressions for these constants.


\subsection{semiclassical trace formula for $\delta P(n)$}
\label{ssectrf}

We are now equipped for doing the approximate Gaussian integrals over the complex 
saddles which go exactly like it is explained in Appendix \ref{secairy}. Hereby for the 
symmetry partners of each generation the phases $\exp[i\tau_0^{(g)}(E=n)]$ in \eq{mug} and 
$\phi_g(E=n)$ in \eq{phigfit}, together with the phases $\pm\exp(\pm i\alpha_g)$ coming 
from the Gaussian integrals along $r$ using \eq{ralpha}, combine to 
$2\cos[n\,\tau_g-3\mu_g\,n^{1/3}+\varphi_g]$, with the constant Maslov index $\varphi_g$ 
given by 
\beq
\varphi_g=\alpha_g-\pi/2\,,
\label{maslov}
\eeq
which is also listed in Tab.\ \ref{tabgen}.
We thus arrive at the semiclassical trace formula for $\delta P(n)$
\beq
\delta P(n) = \sum_{\tau_g>0} A_g(n)
              \cos\left[n\,\tau_g-3\mu_g\,n^{1/3}+\varphi_g\right]\!,
              \qquad (\tau_g \neq 2\pi k, \quad k=1,2,3,\dots)  
\label{dgsc}
\eeq
which is the central result of our paper. Hereby the amplitudes $A_g(n)$ are given by
\beq
A_g(n) = \frac{2}{(4\pi \kappa_g)^{1/2}}\,n^{-5/6}\,
         e^{3\lambda_g\,n^{1/3}}.
\label{ampg}
\eeq
Note that $A_0(n)$ is identical with $P_{as}^{(0)}(n)$ given in \eq{pas0}.
\begin{figure}[h]
\centering
\vspace*{-0.3cm}
\includegraphics[width=9.5cm,angle=0]{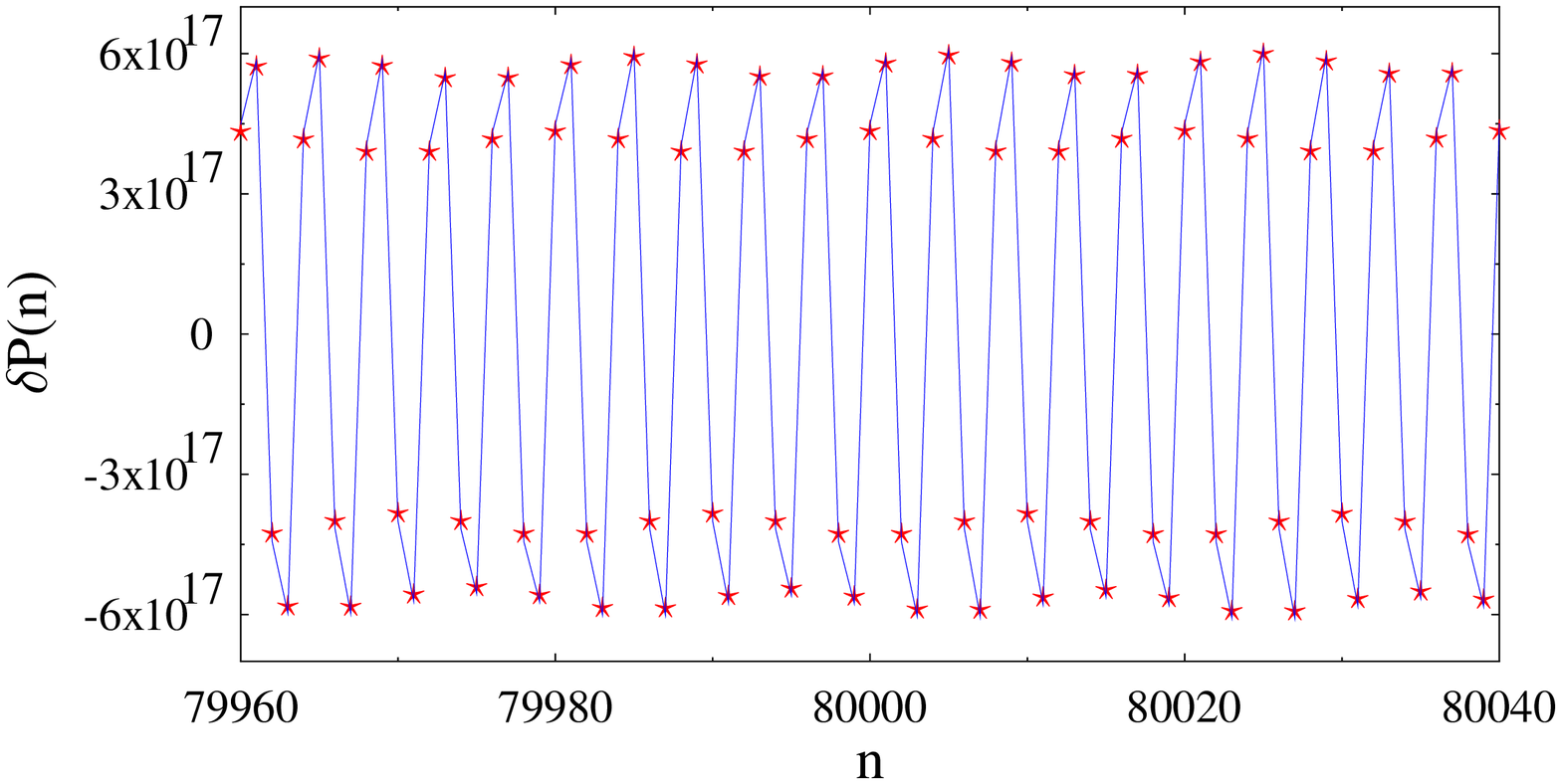}\vspace*{-0.3cm}
\includegraphics[width=9.5cm,angle=0]{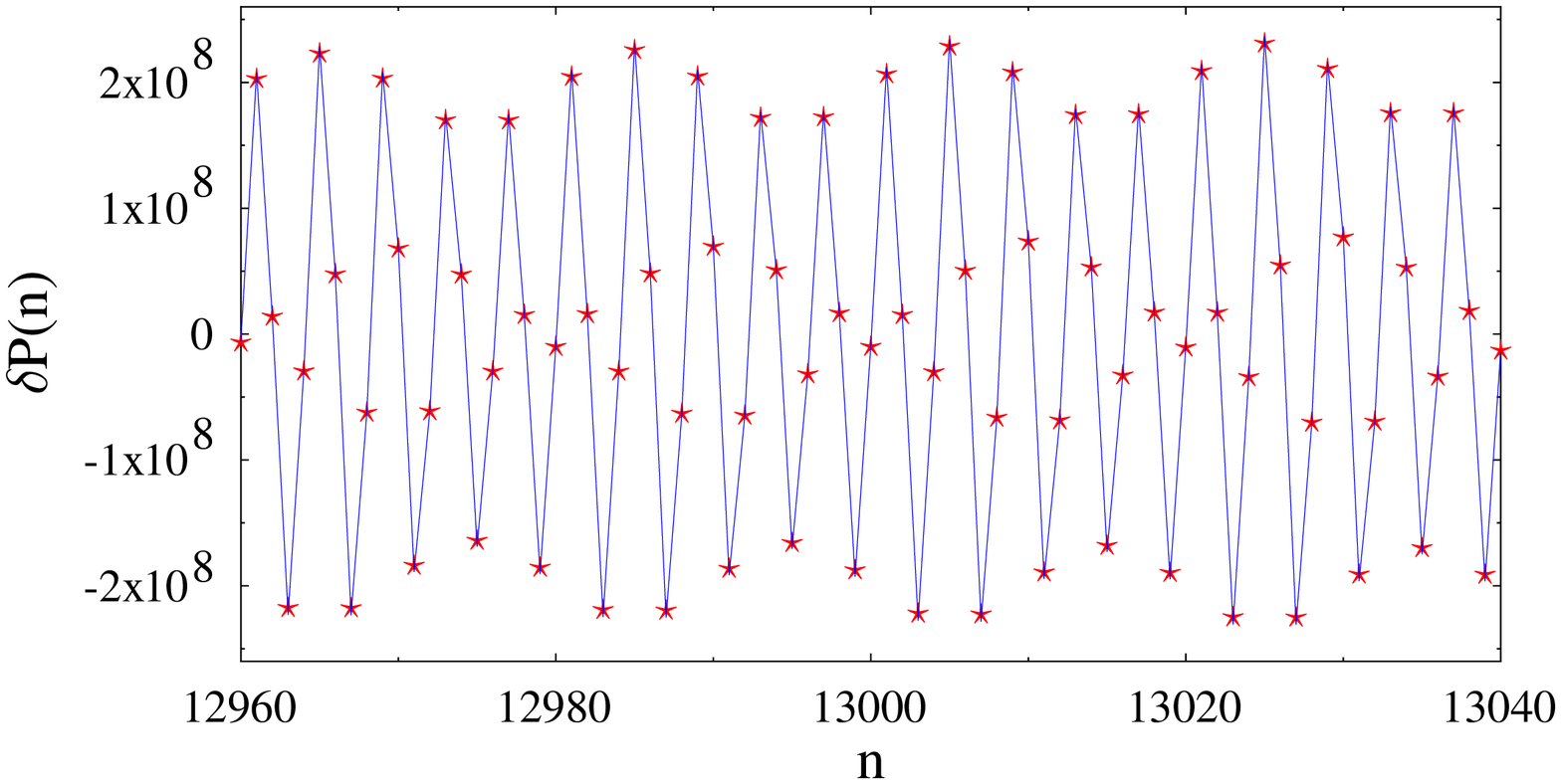}
\vspace*{-0.6cm}
\caption{Result of the trace formula \eq{dgsc}, shown by blue lines, 
versus the exact $\delta P(n)=P(n)-P_{as}(n)$, shown by red stars, in
two regions of large $n$.}
\label{figdp3}
\end{figure} 

\begin{figure}[h]
\centering
\vspace*{-0.2cm}
\includegraphics[width=9.5cm,angle=0]{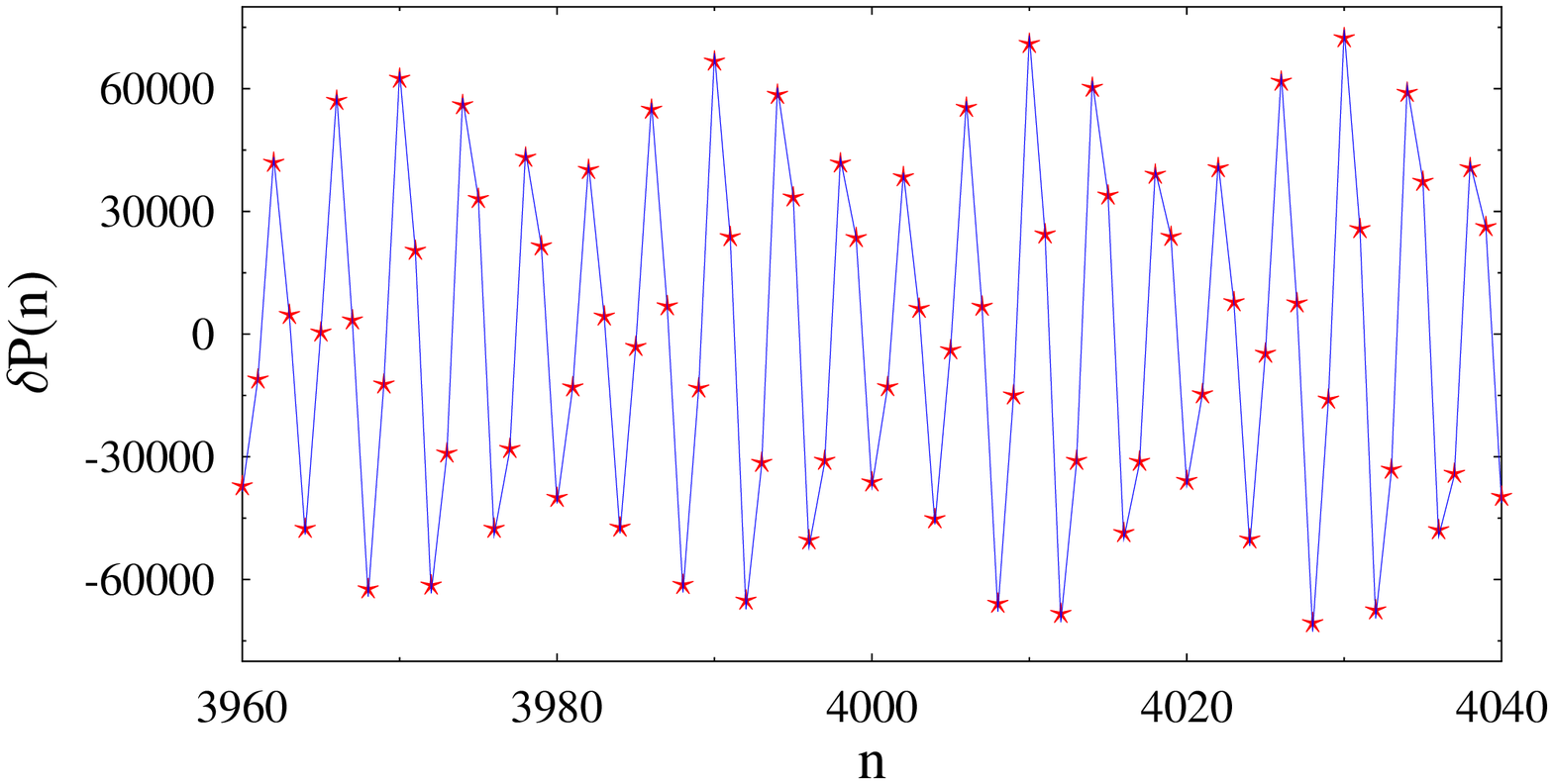}\vspace*{-0.3cm}
\includegraphics[width=9.5cm,angle=0]{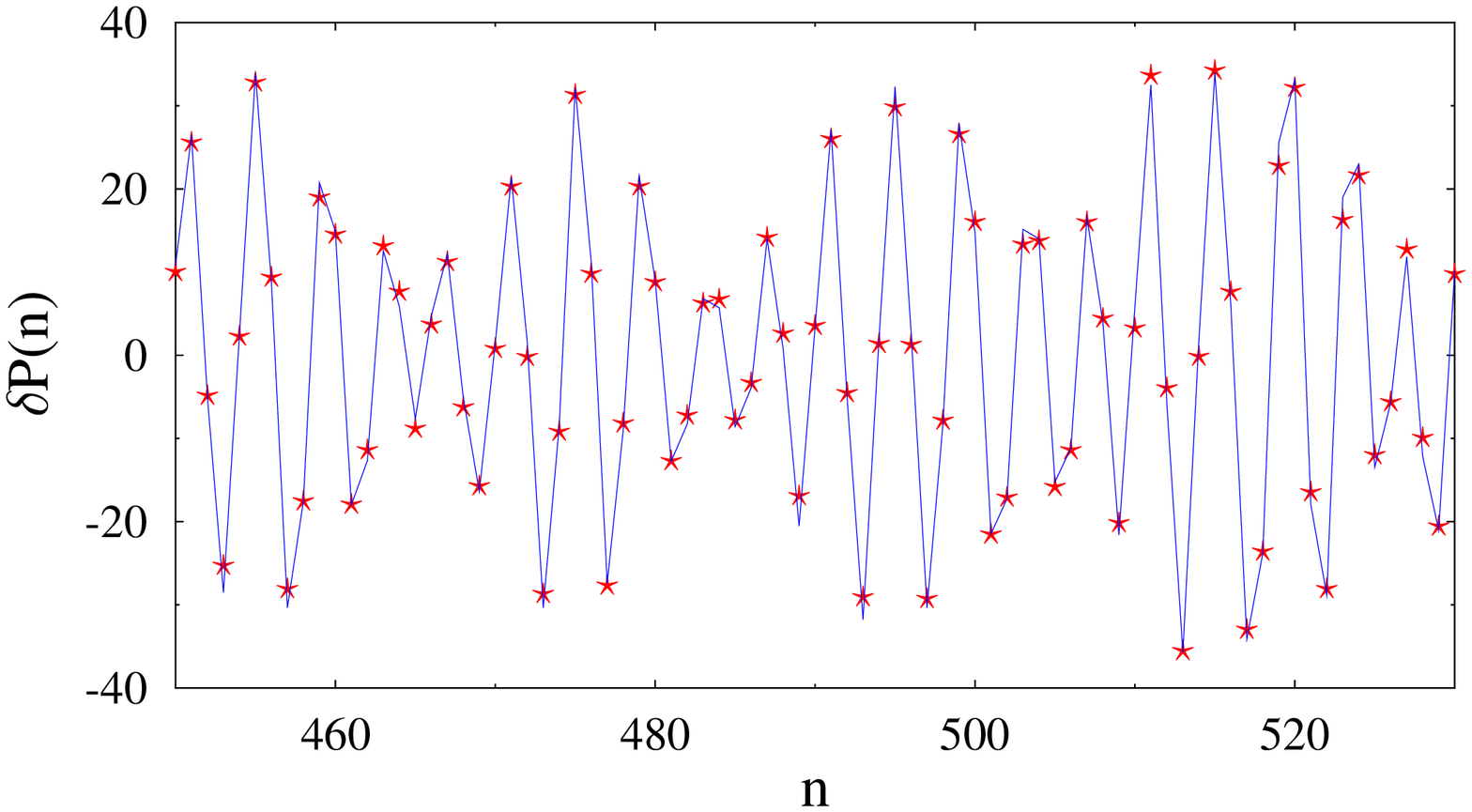}
\vspace*{-0.6cm}
\caption{Result of the trace formula \eq{dgsc}, shown by blue lines, 
versus the exact $\delta P(n)=P(n)-P_{as}(n)$, shown by red stars, in
two regions of small $n$.}
\label{figdp4}
\end{figure}

Figs.\ \ref{figdp3} and \ref{figdp4} show the results of the trace formula \eq{dgsc} by 
blue lines, compared to the exact $\delta P(n)=P(n)-P_{as}(n)$ (red stars)
in four ranges of $n$. The agreement between the two curves is excellent in all regions 
of $n$, the semiclassical results reproducing perfectly both the rapid oscillations of 
the exact $\delta P(n)$ and their beating amplitude.

\vspace*{-0.3cm}

In the calculations for these results, the generations 1-10 have been included. However, 
nothing changes visibly in the results for $n\simg 4000$ if we only include the two
leading generations 1 and 2. While this might be a surprise at first sight, it can be
explained by the values of the constant $\lambda_g$ which regulate the exponential 
growth of the amplitudes $A_g(n)$. These are clearly higher for generations 1 and 2 
than for the others. 

\begin{figure}[h]
\centering
\vspace*{-0.3cm}
\includegraphics[width=11cm,angle=0]{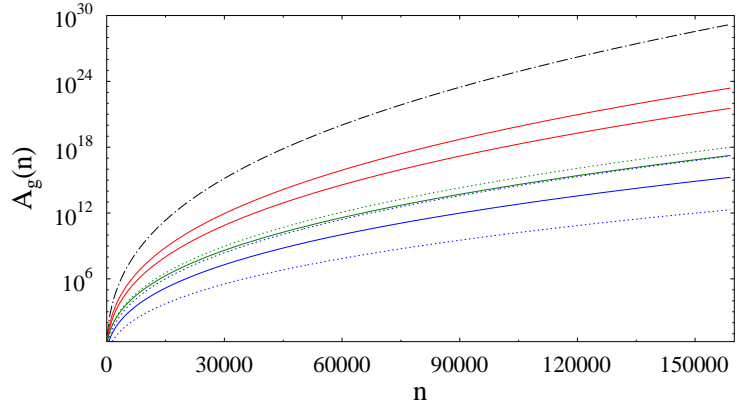}
\vspace*{-0.6cm}
\caption{semiclassical amplitudes $A_g(n)$ for the generations (from top to bottom)
0, 2, 1, 6+7, 3+4+5, 8, 10, 9 (see text for details).}
\label{figamp}
\end{figure}

The relative weights of the generations can be understood from Fig.\ \ref{figamp}, 
where we plot the amplitudes $A_g(n)$ on a logarithmic scale. The long-dashed top line 
gives the amplitude of generation zero, which is identical with $P_{as}^{(0)}(n)$. The 
solid (s) and short-dashed (s-d) lines give, from top to bottom, 
the amplitudes of the generations 2 (s), 1 (s), 6+7 (s-d), 3+4+5 (s), 8 (s-d), 10 (s), 
and 9 (s-d). Note that these amplitudes follow a slightly different ordering than 
that of the generations listed in Tab.\ \ref{tabgen}. 
The amplitudes of the generations 3 and higher are seen to be smaller than those of 
generations 1 and 2 by 2-3 orders of magnitudes for $n\simg 5000$. These in turn are 
smaller than $P_{as}(n)$ by 2-3 orders of magnitude, demonstrating the relative 
smallness of the oscillating part.  

\begin{figure}[h]
\centering
\hspace*{-1.cm}
\vspace*{-2.4cm}
\includegraphics[width=20cm,angle=0]{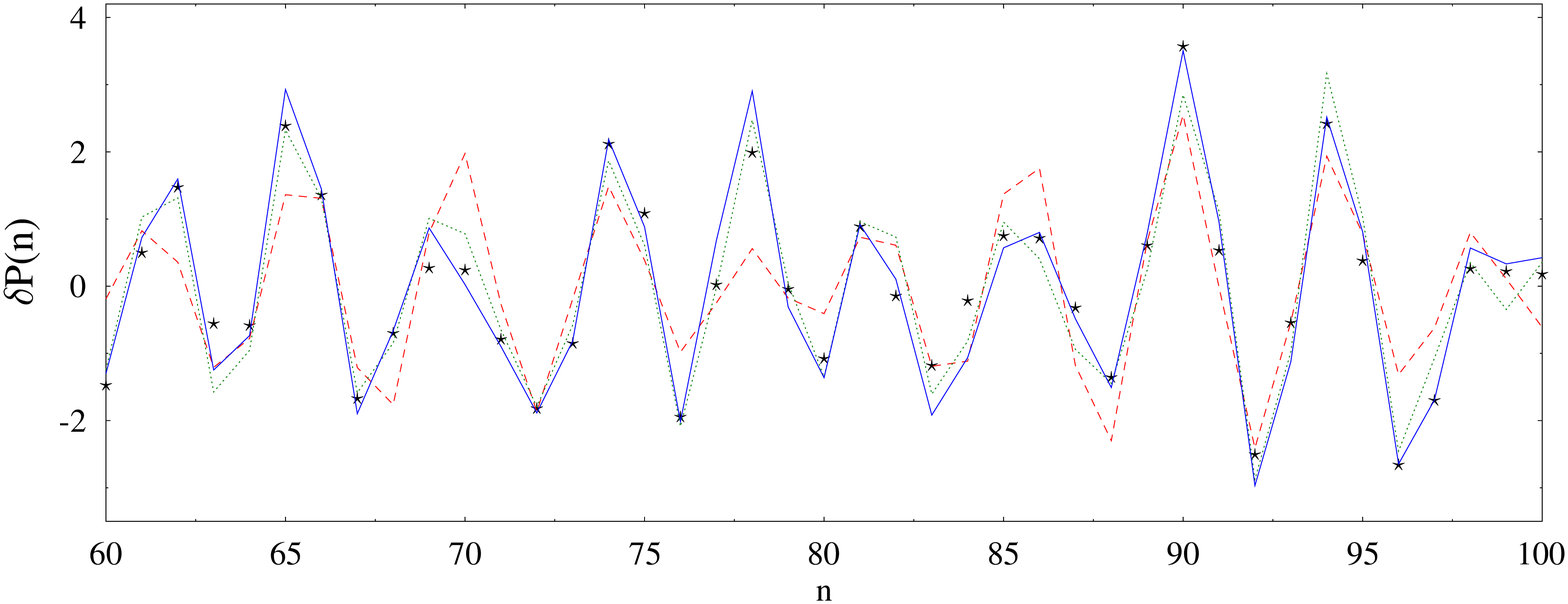}\vspace*{1.2cm}
\caption{Result of trace formula \eq{dgsc} around $n=80$ for increasing numbers of 
generations included. Dashed line (red): generations 1 and 2; dotted line (green): 
generations 1-7; solid line (blue): generations 1-10. The stars (black) show the 
exact $\delta P(n)$.}
\label{figconv}
\end{figure}

The relative importance of the higher generations around $n \sim 80$ can be studied 
in Fig.\ \ref{figconv}. Even here, the generations 1 and 2 produce the essential 
beating part of $\delta P(n)$. The inclusion of higher generations successively 
improves the semiclassical values of $\delta P(n)$, although their contributions 
are rather small and the convergence to the exact values is not as good as for 
$n \simg 500$. Together, these two figures demonstrate the overall rapid 
convergence of the trace formula upon summing over the generations $g$.
\vspace*{-0.4cm}
\vspace*{-0.2cm} 


We conclude that the oscillations in $\delta P(n)$ are dominated everywhere by the 
orbits of generations 1 and 2 with frequencies 4 and 5, which are members of the 
PT (3,4,5). The contributions from all higher generations are practically negligible 
for $n\simg 4000$ and still very small around $n\sim 500$.

In order to understand the beat structure, one must realize that when changing the
variable $E$ to $n$ and studying $\delta P(n)$ as a function of $n$, the roles of the
periods $\tau_g$ and frequencies $(2\pi/\tau_g)$ interchange their roles. The terms 
$\cos(n\tau_g+\dots)$ in \eq{dgsc} have, as functions of $n$, the (approximate) 
periods $2\pi/\tau_g$ and hence frequencies $\tau_g$. In the region where the beat
structure is dominant, the period of the rapid oscillations is roughly that of the 
orbit with the largest amplitude$^\dagger$ (i.e., $\tau_2$ with frequency 4), while the 
beat comes from the difference in their frequencies: the period $\Delta n = 20$ of the 
beat is nothing but one over the inverse frequency difference $1/f_2-1/f_1=1/4-1/5=1/20$. 

\begin{figure}[h]
\centering
\vspace*{-0.4cm}
\includegraphics[width=11.5cm,angle=0]{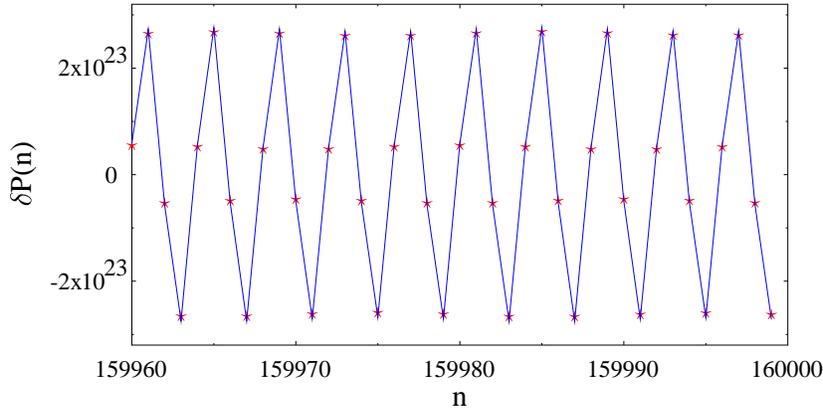}
\vspace*{-0.5cm}
\caption{Result of \eq{dgsc}, shown by the blue line, using only the pair of 
orbits of generation 2 (with $f_2=4$), in the region near $n=160,000$. 
The exact $\delta P(n)=P(n)-P_{as}(n)$ are shown by the red stars.
Note that the beat structure in the exact $\delta P(n)$ has practically
disappeared.
}
\label{fighi}
\end{figure}

For $n\simg 100,000$, the beat structure fades away and the oscillations are 
practically given by the orbits of frequency 4 alone, as shown in Fig.\ \ref{fighi} 
for $n$ near 160,000. The exact values $\delta P(n)$, shown by the stars, exhibit 
practically no more beating amplitude. This is due to the fact that the amplitude of 
generation 2 here is nearly 2 orders of magnitude larger than that of generation 1. 

\vspace*{-0.3cm}

\noindent
------------------------------------------------------------------------------\\
$^\dagger${\small The superposition of two cos functions: $a_1\cos(x_1)+a_2\cos(x_2)$ 
with different periods and similar but not equal amplitudes $a_1>a_2$ yields a 
beat structure, where the rapid oscillation is governed by the period of the 
component with the larger amplitude (i.e.\ $a_1$). (Only for $a_1=a_2$ the rapid 
oscillation has the average period of the two components.)}

\newpage

\noindent
From this result we can give the rate of disappearance of 
the relative oscillations as
\beq
\left|\frac{\delta P(n)}{P_{as}(n)}\right| \sim A_4/A_0 
\sim 2\sqrt{\kappa_0/\kappa_2}\,e^{-3(\lambda_0-\lambda_2)\,n^{1/3}}
\sim 1.9187\,e^{-0.25866\,n^{1/3}}
\quad \hbox{ for } n \rightarrow \infty\,.
\label{dPbyPas}
\eeq

In the region $n\simg 100,000$, the truncated series of three SP corrections in 
\eq{pas1} does not converge fast enough; this shows up in the numerical results by a 
slight asymmetry of the numerical $\delta P(n)$ with respect to its average which 
ought to be zero. We found that a renormalized value ${\tilde c}_3=0.025$ of the 
coefficient $c_3$ given in \eq{c3} makes up for this asymmetry, yielding with 
\eq{pas1} a correct average value $P_{as}(n)$ for all $n$ up to the limit $n=160,000$ 
of our data base.

The above results, demonstrating the absolute dominance of the $f=4$ orbit for
$n\to\infty$, the leading role of the orbits with $f=4$ and $f=5$, creating 
the beating pattern, in the range $4,000 \siml n \siml 100,000$, and the role of 
other orbits with higher frequencies in the range $100 \siml n \siml 4,000$,
are in perfect agreement with the Fourier transforms shown in Fig.\ \ref{figFTP},
which were obtained using Eq.\ \eq{FTZgen} with the corresponding cut-offs in the
$n$ summation.


\section{Summary}
\label{secsum}

We have investigated the number $P(n)$ of partitions of an integer $n$ into sums 
of distinct squares (here called F2 partitions) using semiclassical and quantum 
statistical methods. After some formal definitions in Sec.\ \ref{secbasic} leading 
to the integral representation \eq{intrepP} of $P(n)$, we have in Sec.\ \ref{secasy} 
derived the asymptotic smooth function $P_{as}(n)$ using the stationary-phase method 
like in Refs.\ \cite{tran,primep}, obtaining not only its leading part but also 
higher-order contributions from the so-called saddle-point corrections which lead to 
the correct limit $P(n)/P_{as}(n) \rightarrow 1$ for $n\simg 10,000$. A general method 
for obtaining the latter for arbitrary systems is given in App.\ \ref{secspcor}. The 
smooth part $P_{as}(n)$ is then used to define the oscillating part as $\delta P(n)
= P(n)-P_{as}(n)$.

The oscillations in $\delta P(n)$, with a prominent beat structure for 
$n\le 100,000$, have been analyzed in Sec.\ \ref{secft} through the Fourier 
spectrum of the F2 partition density. The most prominent frequencies were found 
to be integer valued and belong to the lowest Pythagorean triples (PTs) of 
integers $(m,p,q)$ with $m^2+p^2=q^2$, namely (3,4,5) and (5,12,13). Several 
pairs $(p,q)$ of higher PTs $(m,p,q)$ can also clearly be seen in the Fourier 
spectrum. We recall that such triples can only occur in square partitions, since 
Fermat's last theorem \cite{wiles} asserts that only squares of integers may be 
written as sums of two (or more) other squares. It is therefore a particularity
of the F2 partitions that the presence of PTs causes the beating oscillations; 
our Fourier spectra give a clear evidence of this PT dominance. 

In the semiclassical theory of quantum densities of states, oscillations are related 
to sums over periodic functions whose arguments involve the actions of classical 
periodic orbits \cite{gutz,book}. This has been born out quantitatively for the
`dynamical system' of F2 partitions in Sec.\ \ref{sectrf}, in which we have derived 
a semiclassical trace formula for $\delta P(n)$ by exploiting the analytical 
structure of the partition function $Z(\beta)$ \eq{Z2} in the complex plane. 
We have evaluated the exact integral \eq{intrepP} for $P(n)$ approximately by 
deforming the integration contour so as to pass over some selected saddles in the 
complex $\beta$ plane, corresponding to the most prominent Fourier peaks, and by 
locally using stationary-phase integration over these saddles. This leads to the 
trace formula \eq{dgsc} as the central result of our paper.

The numerical results of \eq{dgsc} reproduce the exact $\delta P(n)$ very accurately
all the way from $n\sim 500$ to the upper limit $n=160,000$ of our data base. For
the smallest $n$, up to 10 generations of orbits -- some of them with frequencies
belonging to the PT (5,12,13) -- have been included to reach good agreement. For $n 
\simg 4000$, $\delta P(n)$ is completely determined by the orbits with the frequencies 
4 and 5 belonging to the smallest PT (3,4,5). The period of the rapid oscillations is 
hereby governed by the orbits with the larger amplitude (generation 2, frequency 
$f_2=4$), while the beat period $\Delta n = 20$ is one over their inverse frequency 
difference: $1/f_2-1/f_1=1/4-1/5=1/20$. The contributions from all higher generations 
are negligible for $n\simg 4000$, and still very small around $n\sim 1000$. For $n\simg 
100,000$, the beat structure fades away and the oscillations are given by the orbits 
of frequency 4 alone.

In combination with the Fourier spectra, these results demonstrate an important
role of PTs in establishing the beating oscillations in the F2 partitions $P(n)$. 
In some preliminary statistical studies, we have counted the number $I_N(n)$ of 
pairs of integers belonging to PTs present in distinct square partitions $P_N(n)$ 
restricted to $N$ summands. We found that for $N \simg 20$, the functions $I_N(n)$ 
exhibit the same oscillations as those in the unrestricted F2 partitions $P(n)$ 
and the Fourier spectra of the $P_N(n)$ are practically identical with those shown 
in this paper. Further research along these lines is planned. 

\medskip

R.K.B.\ and M.B.\ are grateful to the IMSc, Chennai, for its hospitality 
and excellent working conditions during the initial stages of this work. 
R.K.B., J.B., and M.V.N.\ sincerely thank Lis Brack-Bernsen for excellent 
hospitality in Matting where an important part of this work was done. Finally 
M.V.N.\ acknowledges the peaceful atmosphere in Apollo Hospitals (OMR), 
Chennai, where parts of the manuscript were completed.



\newpage

\begin{appendix}

\section{Stationary-phase integration: Higher-order contributions}
\label{secspcor}

In this appendix we briefly resume the method of stationary-phase integration including 
saddle-point corrections at higher order for a real saddle, closely following Ref.\ 
\cite{jelovic}.

Consider an integral of the form
\begin{equation}
g(E) =\frac{1}{2\pi i}\int_{\epsilon-i\infty}^{\epsilon+i\infty} {\rm d} \beta\, e^{S(E,\beta)},
\label{apx1}
\end{equation}
where $S(E,\beta)$ is the entropy function defined in \eq{Sinf}. In order to derive
an asymptotic form $g_{as}(E)$ valid for large $E$, we use the method of stationary phase
by expanding the exponent $S(E,\beta)$ around a saddle point (SP). For convenience 
we denote the derivatives of $S$ here by
\begin{equation}
S_n(\beta)=\frac{\partial^n S(E,\beta)}{\partial \beta^n}\,,
\label{apx3}
\end{equation} 
omitting the argument $E$ which for the present development is just a parameter.
We assume that there exists a real SP $\beta_0$ so that
\beq
S_1(\beta_0) = 0\,, \qquad \beta_0 > 0\,.
\eeq
Taylor expanding the entropy $S(E,\beta)$ about $\beta_0$ we have
\begin{equation}
S(E,\beta) =S(E,\beta_0)+S_2(\beta_0)\frac{(\beta-\beta_0)^2}{2}+R(\beta)\,,
\label{apx4}
\end{equation}
where
\begin{equation}
R(\beta)=\sum_{m=3}^{\infty}S_m(\beta_0)\frac{(\beta-\beta_0)^m}{m!}\,.
\label{apx5}
\end{equation}
Hence we obtain an asymptotic form
\begin{equation}
g_{as}(E) = \frac{e^{S(E,\beta_0)}}{2\pi i} \int_{\epsilon-i\infty}^{\epsilon+i\infty} {\rm d}\beta\,  
           e^{S_2(\beta_0)\frac{(\beta-\beta_0)^2}{2!}}\,e^{R(\beta)}.
\label{apx6}
\end{equation}
Without loss of generality we can choose $\epsilon=\beta_0$ and define the variable 
\begin{equation}
u=(\beta-\beta_0)/i\,.
\label{apx7}
\end{equation} 
Expanding the exponential of $R$ under the integral we find
\begin{equation}
g_{as}(E) = \frac{e^{S(\beta_0)}}{2\pi} \int_{-\infty}^{\infty} {\rm d}u\,  
           e^{-\frac12 S_2(\beta_0)\,u^2}
           \!\left[1+R(u)+\frac{R^2(u)}{2!}+\frac{R^3(u)}{3!}+\cdots\right].
\label{apx8}
\end{equation}
This leads to Gaussian integrals over $u$ such that only even powers of $u$ contribute, 
leaving $g(E)$ real as it should be. Collecting the even powers of $u$, we get
\begin{eqnarray}
g_{as}(E) = \frac{e^{S(\beta_0)}}{2\pi} \int_{-\infty}^{\infty} {\rm d}u\,  
       e^{-\frac12 S_2(\beta_0)\, u^2}
       \!\left[1+\sum_{m=2}^\infty (-1)^m u^{2m}\sum_{\{k\}}
       \prod_{i=1}^k\left(\frac{S_{n_i}}{n_i!}\right)^{m_i}\!
       \left(\frac{1}{m_i!}\right)\right]\!,
\label{apx9}
\end{eqnarray}
where the sum over $\{k\}$ implies the following constraints:
\begin{equation}
2m=m_1n_1+m_2n_2+\cdots + m_kn_k\,, \qquad n_i\ge 3\,, \qquad k\ge 1\,,
\label{apx10}
\end{equation}
which is the allowed number of partitions of $2m$ into $k$ parts with 
repetitions allowed through the power $m_i$. All such partitions 
contribute at order $2m$ in $u$. The integration is now straightforward, 
since the basic integrals needed are simply
\beq
\int_{-\infty}^\infty {\rm d}u\,e^{-au^2}u^{2m} 
     = \frac{(2m-1)!!}{2^ma^m}\sqrt{\pi/a}\,.
\eeq
With this we obtain a result that formally contains SP crrections to all orders:
\begin{eqnarray}
g_{as}(E) = \frac{e^{S(\beta_0)}}{\sqrt{2\pi S_2(\beta_0)}}\,
       \left[1+\sum_{m=2}^\infty (-1)^m \frac{(2m-1)!!}{ S_2^m}\sum_{\{k\}}
       \prod_{i=1}^k\left(\frac{S_{n_i}}{n_i!}\right)^{m_i}
       \left(\frac{1}{m_i!}\right)\right]\!.
\label{apx11}
\end{eqnarray}
This is a series with $1/S_2^2$ as the expansion parameter. Even though
this result is written to all orders, we note that it is an asymptotic
series obtained by Taylor expansion of the entropy around a point $\beta=
\beta_0$. Hence, depending on the system under consideration, truncation
of the series must be handled with care. 

We now use Eq.\ (\ref{apx11}) to obtain the SP corrections to the asymptotic
F2 partition density given in \eq{g0as}. To leading order we have
\begin{equation}
S(E,\beta) = \beta E+ \frac{D}{\sqrt{\beta}}-\frac{1}{2}\ln(2)\,,
\label{apx12}
\end{equation}
and at the real SP $\beta_0$ we have
\begin{equation}
E=\frac{D}{2\beta_0^{3/2}}\,, \qquad \beta_0=\left(\frac{D}{2E}\right)^{2/3}.
\label{apx13}
\end{equation}
The derivatives of $S(E,\beta_0)$ are given by
\begin{equation}
S_n=(-1)^n\frac{(2n-1)!!D}{2^n\beta_0^{(2n+1)/2}}\,, \qquad n\ge 2\,,
\label{apx14}
\end{equation}
with the expansion parameter
\begin{equation}
S_2=\frac{3!!D}{2^2\beta_0^{5/2}}\,.
\label{apx15}
\end{equation}

For our purposes it is more convenient to have $\beta_0$ (or $E^{-2/3}$) as the 
expansion parameter rather than $1/S_2$. To achieve this order by order, consider
a typical term in the expansion given in Eq.(\ref{apx11}) of the form
\begin{equation}
\frac{S_{n_1}S_{n_2}...S_{n_k}}{S_2^m}
     \propto (\beta_0)^{5m/2 -(2n_1+1)/2-(2n_2+1)/2-...-(2n_k+1)} = (\beta_0)^M.
\label{apx16}
\end{equation}
For counting and collecting the powers in $\beta_0$ it is convenient to put 
$m_i=1$ and allow repetitions of $n_i$. Imposing the constraints \eq{apx10} we have
\begin{equation}
M = \frac{5m}{2}-(n_1+n_2+...+n_k)-\frac{k}{2}=\frac{m-k}{2}\,, \qquad n_i\ge 3\,,
\label{apx17}
\end{equation}
since
\beq
n_1+n_2+...+n_k=2m\,.
\eeq
Here $M$ is the power of $\beta_0$ in the series and $k$ is the number of 
terms in the product that contribute to a given power $M$ which may be 
more than one. Since $m\ge 2$ and $k\ge 1$, the lowest power is $M=1/2$. 
In Tab.\ \ref{tabsp} we give the relevant contributions up to $M=1/2,1,3/2$. 

\begin{table}[h]
\centering
\begin{tabular}{|ccccc|}\hline
M=(m-k)/2 & m & k & $n_i$ & Terms\\\hline
1/2 & 2 & 1 & 4 & $S_4/S_2^2$\\
1/2 & 3 & 2 & 3,3 & $S_3^2/S_2^3$\\\hline
1 & 3 & 1 & 6 & $S_6/S_2^3$\\
1 & 4 & 2 & 4,4 & $S_4^2/S_2^4$\\
1 & 4 & 2 & 5,3 & $S_3S_5/S_2^4$\\
1 & 5 & 3 & 4,3,3 & $S_4S_3^2/S_2^5$\\
1 & 6 & 4 & 3,3,3,3 & $S_3^4/S_2^6$\\\hline
3/2 & 4 & 1 & 8 & $S_8/S_2^4$\\
3/2 & 5 & 2 & 7,3 & $S_7S_3/S_2^5$\\
3/2 & 5 & 2 & 6,4 & $S_6S_4/S_2^5$\\
3/2 & 5 & 2 & 5,5 & $S_5^2/S_2^5$\\
3/2 & 6 & 3 & 6,3,3 & $S_6S_3^2/S_2^6$\\
3/2 & 6 & 3 & 5,4,3 & $S_5S_4S_3/S_2^6$\\
3/2 & 6 & 3 & 4,4,4 & $S_4^3/S_2^6$\\
3/2 & 7 & 4 & 5,3,3,3 & $S_5S_3^3/S_2^7$\\
3/2 & 7 & 4 & 4,4,3,3 & $S_4^2S_3^2/S_2^7$\\
3/2 & 8 & 5 & 4,3,3,3,3 & $S_4S_3^4/S_2^8$\\
3/2 & 9 & 6 & 3,3,3,3,3,3 & $S_3^6/S_2^9$\\\hline
\end{tabular}
\vspace*{0.2cm}
\caption{Contributing terms at each order in powers of $\beta_0$. Note
that $n_i\ge 3$ and hence the saturation occurs when all or most of
$S_n$ are equal to 3.}
\label{tabsp}
\end{table}
\noindent
The table lists all the terms which contribute at a particular order in
$\beta_0$. Putting in the numerical factors from integration as given 
in Eq.\ \eq{apx11} and collecting terms at each order, we obtain
\begin{eqnarray}
& &g_{as}(E) =\frac{e^{S(\beta_0)}}{\sqrt{2\pi S^{(2)}}}
\{1+\left[\frac{3!!(S_4/4!)}{(S_2)^2}-\frac{5!!(S_3/3!)^2}{2!(S_2)^3}\right]
\nonumber\\
&+&\left[-\frac{5!!(S_6/6!)}{(S_2)^3}
+\frac{7!!((S_3S_5/3!5!)+(S_4/4!)^2/2!)}{(S_2)^4}
-\frac{9!!(S_3/3!)^2(S_4/4!)/2!}{(S_2)^5}
+\frac{11!!(S_3/3!)^4/4!}{(S_2)^6}\right]\nonumber\\
&+&\frac{7!!S_8}{(S_2)^4}-\frac{9!!((S_7S_3/3!7!)+(S_6S_4/6!4!)
+(S_5/5!)^2/2!)}{(S_2)^5}\nonumber\\
&+&\frac{11!!((S_6/6!)(S_3/3!)^2/2!+(S_5S_4S_3/5!4!3!)+(S_4/4!)^3/3!)}
{(S_2)^6}\nonumber\\
&-&\frac{13!!((S_5/5!)(S_3/3!)^3/3!+(S_4S_3/4!3!)^2/2!2!)}
{(S_2)^7}\nonumber\\
&+&\frac{15!!((S_4/4!)(S_3/3!)^4/4!)}
{(S_2)^8}\nonumber\\
&-&\frac{17!!((S_3/3!)^6/6!)}
{(S_2)^9}-\cdots \}.
\label{apx18}
\end{eqnarray}
Substituting for $S_n$ from Eq.(\ref{apx14}), we finally get the desired series 
of contributions
\begin{equation}
g_{as}(E)=
\frac{e^{S(\beta_0)}}{\sqrt{2\pi S^{(2)}}}\left[
1-C_1\left(\frac{\sqrt{\beta_0}}{D}\right)
-C_2\left(\frac{\sqrt{\beta_0}}{D} \right)^2
-C_3\left(\frac{\sqrt{\beta_0}}{D}\right)^3 
-\cdots\right]\!,
\label{apx19}
\end{equation}
where 
\begin{equation}
C_1=\frac{5}{18}=\frac{5}{2\times3^2}, \qquad
C_2=\frac{35}{648}=\frac{5\times 7}{2^3\times3^4}, \qquad
C_3=\frac{665}{34992}=\frac{5\times7\times19}{2^4\times3^7}.
\label{apx20}
\end{equation}
Furthermore
\begin{equation}
\frac{\sqrt{\beta_0}}{D}=\left[\frac{1}{2D^2E}\right]^{1/3}, \qquad D=0.678093895\,.
\end{equation}
Substituting for $\beta_0$ we finally obtain 
\begin{equation}
g_{as}(E)=
\frac{e^{S(\beta_0)}}{\sqrt{2\pi S^{(2)}}}\left[
1-c_1E^{-1/3}-c_2E^{-2/3}-c_3E^{-1} -\cdots\right],
\label{apx22}
\end{equation}
with
\beq
c_1=\frac{C_1}{(2D^2)^{1/3}}=0.285645648\,, \quad
c_2=\frac{C_2}{(2D^2)^{2/3}}=0.057115405\,,
\label{c12}
\eeq
and
\beq
c_3=\frac{C_3}{(2D^2)^{3/3}}=0.020665371\,.
\label{c3}
\eeq

\section{Asymptotic evaluation of the Airy function}
\label{secairy}

We derive here the asymptotic expressions of the real-valued Airy function Ai\,$(z)$, 
both for large positive and negative real $z$, using the stationary-phase method, 
as an illustration of the method used in the main text. In Ref.\ \cite{BPD} this 
was done for integrals over the Airy function; here we do it for the Airy function 
itself. Its complex integral representation is given by
\be
{\rm Ai}\,(z) = \frac{1}{2\pi i} \int_C d\beta \, e^{S(z,\beta)}, \qquad \beta=x+iy\,,
\label{Aiz}
\ee
with
\be S(z,\beta) = -z\beta+\frac13\,\beta^3\,.
\label{Sairy}
\ee
C is a contour along the imaginary $\beta$ axis, i.e. from $y=-\infty$ to $y=+\infty$, 
at a finite distance $x=\epsilon>0$ (see also Fig. 1 in Ref.\ \cite{BPD}). Let us 
split the function $S(z,\beta)$ in its real and imaginary part:
\be
S(z,\beta) = X(x,y) + i\,Y(x,y)\,,
\ee
so that
\be
X(x,y) = -zx+\frac13\,x^3-xy^2\,, \qquad Y(x,y) = -zy-\frac13\,y^3+x^2y\,.
\ee
(We ignore the argument $z$ which is a parameter in the real functions $X$ and $Y$.) 
It is easy to see that the Cauchy-Riemann (CR) conditions are fulfilled for $X$ and $Y$:
\be
\frac{\partial X}{\partial x} = \frac{\partial Y}{\partial y} = -z+x^2-y^2,
\label{CR1}
\ee
and
\be
\frac{\partial X}{\partial y} = - \frac{\partial Y}{\partial x} = -2xy\,.
\label{CR2}
\ee
Therefore $S(z,\beta)$ is analytic in the whole complex $\beta$ plane; it has no poles. 
[The shaded areas in Fig.\ 1 of \cite{BPD} are those in which Re $\beta^3<0$ so that 
the integrand of \eq{Aiz} vanishes at both ends of the contour C.]

Like in \cite{BPD}, we evaluate (\ref{Aiz}) approximately in the stationary-phase 
approximation. To that purpose we solve the saddle-point (SP) equation and look for
solutions in the complex plane.
\be
\left. \frac{\partial S}{\partial\beta}\right|_{\beta_0} 
        = -z + \beta_0^2 = 0 \quad \Rightarrow \quad \beta_0^2 = z\,.
\label{sp}
\ee
For $z>0$, we have one real solution $\beta_0$:
\be
\beta_0 = \sqrt{z} \quad \Rightarrow \quad x_0=\sqrt{z}\,, \;\; y_0=0\,.
\ee
(We can neglect the negative root of $z$ for the reason given at the end of Subsect. 
1 below.) For $z<0$, we have a pair of imaginary solutions $\beta_{1,2}$:
\be
\beta_{1,2} = \pm \,ai \quad \Rightarrow \quad x_{1,2} = 0, \;\; y_1=a, \;y_2 = -a\,, 
              \qquad a = +\sqrt{|z|}\,.
\ee

We now approximate the integration by the stationary-phase method, obtaining thereby 
the asymptotically leading contributions from the regions near the saddle points. For 
$z>0$ we use the real saddle point $\beta_0$, and for $z<0$ the two complex-conjugate 
saddle points $\beta_{1,2}$. This is also shown and explained by Balasz {\it et al.} 
\cite{BPD}. These authors investigated integrals over the Airy function. Our present 
case, the Airy function itself corresponds to taking $n=-1$ in their treatment. 
Although they exclude negative values of $n$, their results apply also for $n=-1$; 
the complex saddles then lie on the imaginary $t$ axis as shown below. 

\subsection{Case $z>0$: the exponential tail of Ai\,$(z$)}

We first look at $z>0$ and derive the asymptotic expression of Ai\,$(z)$ for $z\gg 0$, 
which is found from SP integration over the real saddle at $\beta_0=x_0=z$. We find 
that the curvature of Re $S=X(x_0,y)$ is negative in the $y$ direction:
\be
\left. \frac{\partial^2 X}{{\partial y}^2}\right|_{y_0=0} = -x_0 = -z\,,
\ee
(and, due to the CR conditions, positive in the $x$ direction), so that a straight-line 
contour along the imaginary axis with $x_0=z$ will lead to a maximum of Re $S$ at 
$y_0=0$. We thus choose the contour C$_0$ as:
\be
{\rm C}_0: \qquad \beta= \sqrt{z} + it\,, \qquad t\in (-\infty,+\infty)\,.
\label{C0}
\ee
Expanding $S(z,\beta)$ along this contour up to order $t^2$, we get
\be
S_0(z,\beta) = -z(\sqrt{z}+it)+\frac13\,(\sqrt{z}+it)^3 
             = -\frac23\,z^{3/2}-\sqrt{z}\,t^2 + \dots,
\ee
and the integral (\ref{Aiz}) yields the result
\be
{\rm Ai}(z) \sim \frac{1}{2\pi i}\,e^{-\frac23\,z^{3/2}}\!\int_{-\infty}^{+\infty} idt\,
                 e^{-\sqrt{z}\,t^2}
              =  \frac{1}{2\sqrt{\pi}z^{1/4}}\,e^{-\frac23\,z^{3/2}}\,,  \qquad (z\gg 0)
\label{aitail}
\ee
which is exactly the leading term of Eq.\ 10.4.59 in \cite{abro}. 

A note concerning the sign of $\sqrt{z}$. In principle, we have two real roots of 
(\ref{sp}) for $z>0$: $\beta_0=\pm \sqrt{z}$. However, if we integrate over the saddle 
$x_0=-\sqrt{z}$, $y_0=0$ as above, we obtain a result like (\ref{aitail}) but with the 
diverging exponential $e^{\frac23\,z^{3/2}}$. This corresponds to the associated Airy 
function Bi$(z)$ \cite{abro} which has the same integral representation as (\ref{Aiz}) 
with an appropriately chosen contour C.

\subsection{Case $z<0$: the oscillations of Ai\,$(z$)}

We now want to integrate along paths that go over the imaginary saddles $\beta_{1,2}$, 
in order to find the asymptotic oscillations of Ai\,$(z)$ for $z \ll 0$. To that 
purpose, let us have a look at the landscape of Re $S(\beta)$ given by Eq.\ \eq{Sairy}.
\begin{figure}[H]
\centering
\includegraphics[width=0.4\columnwidth,clip=true,angle=-90]{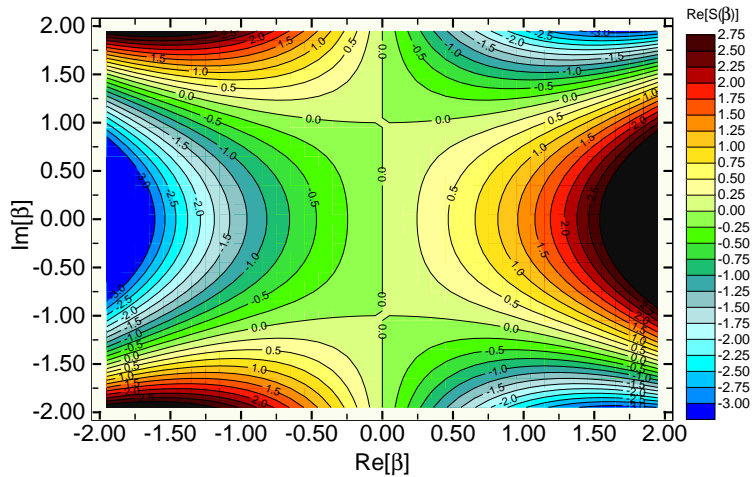}
\caption{Surface plot of Re $S(\beta)$ \eq{Sairy} of the Airy function in the complex 
$\beta$ plane for $z=-1$.}
\label{fig1} 
\end{figure} 
Figure \ref{fig1} shows a plot of Re $S(\beta)$ in the complex $\beta$ plane, taken for 
$z=-1$ (so that $a=1$). We clearly see the two saddles at Im $\beta_{1,2} = y_{1,2} = 
\pm a = \pm 1$ with Re $\beta_{1,2}=0$. We also see that Re $S$ is zero along the imaginary
axis and locally (at $y_{1,2}=\pm a$) in $x$ direction around $x_{1,2}=0$, so that passing 
over the saddles in $y$ or $x$ direction would lead to a zero result. Instead, we have 
(similarly to \cite{BPD}) to pass over the lower saddle (at $y_2=-a$) from lower right 
to upper left (SE to NW), and over the upper saddle (at $y_1=+a$) from lower left to 
upper right (SW to NE), connecting the two paths smoothly to the regions to the right 
of the imaginary axis where the integrand vanishes for $y \to \pm \infty$.

In order to find the directions of steepest descent (or ascent), we parameterize the 
contour over the saddles locally as straight lines and write them in polar coordinates 
$(r,\alpha)$
\be
x=r\cos\alpha\,, \quad y=r\sin\alpha\,.
\ee
Let us start at the upper saddle at $y_1=+a$, $x_1=0$. We define a straight-line contour 
C$_1$ as:
\be
{\rm C}_1: \qquad \beta = ia + r\,e^{i\alpha}\,, \qquad r \in (-\infty,+\infty)\,.
\ee
Along this path, the function $S(z,\beta)$ becomes (noting that $-z=a^2$)
\be
S_1(z,\beta) = a^2(ia+r\,e^{i\alpha})+\frac13(ia+r\,e^{i\alpha})^3
             = \frac23ia^3 +iar^2\,e^{2i\alpha}+\frac13r^3\,e^{3i\alpha}.
\ee
The real part of $S_1$ then is
\be
{\rm Re}\, S_1(r,\alpha) = X_1(r,\alpha) = -ar^2\sin(2\alpha)+\frac13r^3\cos(3\alpha)\,.
\ee
The curvature in the $r$ direction, taken at $r=0$, is
\be
K_1(\alpha) = \left. \frac{\partial^2 X_1}{{\partial r}^2}\right|_{r=0} 
            = -2a\,\sin(2\alpha)\,.
\ee
This has a minimum at $\alpha=\pi/4$ and a maximum at $\alpha=3\pi/4$. Thus the path of 
steepest ascent over this saddle is locally a straight line at the angle $\alpha=\pi/4$,
with $K_1(\pi/4)=-2a$, so that $X_1 (r,\alpha\!=\!\pi/4)$ has a maximum at $r=0$. Along 
this direction the action becomes, up to order $r^2$, 
\be
S_1(z\!=\!-a^2,r) \simeq \frac23\,ia^3 -ar^2.
\ee
Doing the integral over $r$ (taken from $-\infty$ to $+\infty$ to complete the Gauss 
integral) -- and not forgetting that $d\beta=e^{i\alpha}dr$ -- we get the following 
contribution to (\ref{Aiz}):
\be
\frac{1}{2\pi i} \int_{C_1} d\beta \, e^{S(z,\beta)} 
            \sim \frac{-i}{2\sqrt{\pi a}}\,e^{\frac23\,ia^3+i\pi/4}.
\label{intC1}
\ee

At the lower saddle ($y_2=-a$, $x_2=0$), we do exactly the same, defining the contour 
C$_2$ through it:
\be
{\rm C}_2: \qquad \beta = -ia + r\,e^{i\alpha}\,, \qquad r \in (-\infty,+\infty)\,.
\ee
Along this path, the function $S(z,\beta)$ becomes
\be
S_2(z\!=\!-a^2,r) = -\frac23ia^3 -iar^2\,e^{2i\alpha}+\frac13r^3\,e^{3i\alpha}.
\ee
Here the curvature at $r=0$ becomes
\be
K_2(\alpha) = 2a\sin(2\alpha)\,,
\ee
which is maximum at $\alpha=\pi/4$ and minimum at $\alpha=3\pi/4$. Thus we have to go 
over this saddle in the direction $\alpha=3\pi/4$. Proceeding as above, we obtain
\be
\frac{1}{2\pi i} \int_{C_2} d\beta \, e^{S(z,\beta)} 
            \sim \frac{-i}{2\sqrt{\pi a}}\,e^{-\frac23\,ia^3+i3\pi/4}.
\label{intC2}
\ee
Adding the contributions (\ref{intC1}) and (\ref{intC2}), we obtain the following result:
\be
{\rm Ai}(z) \sim  \frac{1}{\sqrt{\pi}\,|z|^{1/4}}\,
                  \sin\left(\frac23|z|^{3/2}+\pi/4\right),  \qquad (z\ll 0)
\label{aiosc}
\ee
which is exactly the leading term of the asymptotic expression 10.4.60 in \cite{abro}.

This exercise demonstrates how the use of the stationary-phase integration over complex 
saddles can yield asymptotic expressions for oscillating functions.

\end{appendix}


\bigskip

\begin{thebibliography}{99}
 
\bibitem{tran}    Muoi N. Tran {\it et al.}, Ann.\ Phys.\ (N.Y.) {\bf 311}, 
                  204 (2004).

\bibitem{hr}      G. H. Hardy and S. Ramanujan, Proc.\ London Math.\ Soc.\ 2, 
                  XVII:75 (1918).

\bibitem{apostol} T. M. Apostol: {\it Introduction to Analytic Number Theory}
                  (Berlin, Springer International Edition, 1989), p.\ 304.

\bibitem{wiles}   D. Castelvecchi: {\it Fermat's Last Theorem earns Andrew Wiles 
                  the Abel Prize}, Nature {\bf 531}, 287 (17 March 2016). 

\bibitem{gutz}    M. C. Gutzwiller, J. Math.\ Phys.\ {\bf 12}, 343 (1971);\\  
                  R. Balian and C. Bloch, Ann.\ Phys.\ (N.Y.) {\bf 69}, 76 (1972);\\ 
                  M. V. Berry and M. Tabor, Proc. R. Soc.\ Lond.\ A {\bf 349}, 101 
                  (1976). 

\bibitem{book}    M. Brack and R. K. Bhaduri: {\it semiclassical Physics} 
                  (Bolder, Westview Press, 2003).

\bibitem{oeis}    {\it The On-Line Encyclopedia of Integer Sequences (OEIS)}: 
                  $<$http://oeis.org/$>$.  

\bibitem{schiff}  L. L. Schiff: {\it Quantum Mechanics}, 3rd edition (McGraw-Hill 1968).

\bibitem{primep}  J. Bartel, R. K. Bhaduri, M. Brack, and M. V. N. Murthy,
                  Phys.\ Rev.\ E {\bf 95}, 052108 (2017). 

\bibitem{abro}    M. Abramowitz and I. A. Stegun: {\it Handbook of Mathematical 
                  Functions}, 9th printing (New York, Dover, 1972).

\bibitem{gr}      I. S. Gradshteyn and I. M. Ryzhik: {\it Table of Integrals, 
                  Series, and Products} 
                  (New York, Academic Press, 5th edition, 1994).

\bibitem{jelovic} M. R. Hoare, J.\ Chem.\ Phys.\ {\bf 52}, 5695 (1970); \\
                  A. Jelovic, Phys.\ Rev.\ C {\bf 76}, 017301 (2007).

\bibitem{BPD}     N. L. Balazs, H. C. Pauli and O. B. Dabbousi, 
                  Math. of Comp., Vol. {\bf 33} No. 145 (1979) pp. 353-358. 

\end{thebibliography}
\end{document}